\documentstyle[citeauthoryear]{lmamult}
\begin{document}

\def\la{\mathrel{\mathchoice {\vcenter{\offinterlineskip\halign{\hfil
$\displaystyle##$\hfil\cr<\cr\sim\cr}}}
{\vcenter{\offinterlineskip\halign{\hfil$\textstyle##$\hfil\cr
<\cr\sim\cr}}}
{\vcenter{\offinterlineskip\halign{\hfil$\scriptstyle##$\hfil\cr
<\cr\sim\cr}}}
{\vcenter{\offinterlineskip\halign{\hfil$\scriptscriptstyle##$\hfil\cr
<\cr\sim\cr}}}}}
\def\ga{\mathrel{\mathchoice {\vcenter{\offinterlineskip\halign{\hfil
$\displaystyle##$\hfil\cr>\cr\sim\cr}}}
{\vcenter{\offinterlineskip\halign{\hfil$\textstyle##$\hfil\cr
>\cr\sim\cr}}}
{\vcenter{\offinterlineskip\halign{\hfil$\scriptstyle##$\hfil\cr
>\cr\sim\cr}}}
{\vcenter{\offinterlineskip\halign{\hfil$\scriptscriptstyle##$\hfil\cr
>\cr\sim\cr}}}}}

\title{The Gold Effect: Odyssey of Scientific Research}
\author{Wolfgang Kundt}

\institute{Institut f\"ur Astrophysik der Universit\"at, D-53121 Bonn}

\maketitle

\noindent{\bf Abstract:}\\
Seventy-nine physical problems are listed and explained with whose 
proposed solutions I do not agree. Such disagreements -- even though 
some of them may simply reflect our preliminary insight into the laws 
of nature -- have occasionally caused deplorable damage to personal 
relationships.

\section{Introduction}

Raymond Lyttleton (1981) has explained under the term `Gold Effect' what had been emphasized to him by Thomas Gold: that a mere unqualified belief can occasionally be converted into a generally accepted scientific theory, through the screening action of 
refereed literature, of meetings organized by scientific organizing committees, and through the distribution of  funds controlled by `club' opinions. In the (last) chapter `Cargo Cult Science' of his book `Surely You're Joking Mr. Feynman' (1985), Richard 
Feynman gives lucid examples of same phenomenon. Quite generally, it occurs to me that any physical result -- in astrophysics, geophysics, biophysics, or else -- which has not benefitted from experimental tests has a high chance of being wrong: physics is 
not all that easy.

In the table called `ALTERNATIVES' at the end of this essay, I summarize seventy-nine examples of where I have been unable to follow established wisdom, ordered (more or less) historically but collected into groups of related topics. The (last) column 
superscribed `Year' lists the year (of this century) when my first and/or `best' writeup on that problem appeared in print -- which occasionally happened several years after the first glimpse of scepticism. There were cases when submitted manuscripts got lost, 
were repeatedly rejected because journals felt `unable to publish' its contents, where I was not invited to topical meetings, and even where I was sent home from a meeting on the day of my arrival. I may well have erred on various of the 79 alternatives 
(because no human mind is always right): please let me know. This essay will rush through all of them, state the problem, list references pro and con, and add a little anecdote whenever considered appropriate. A representative figure, or set of formulae, 
will go with almost every item and be explained in the text.

When I decided to study physics (in 1950, in Hamburg), it was my understanding that physics was an objective science whose frontier domains simply had to be extended and/or worked out, in order to one day yield a complete `picture' of nature, or at least a 
unique basis for such. It was not at all my intention to rebel against existing `knowledge'. In order not to leave the reader under the (false?) impression that I was hunting for `alternatives', I will comment in section 3 on a few further publications in which I have tried to help exploring the `grand design'.

\section{The seventy-nine cases}

{\bf 1. Neutron-star crusts} are of interest because they have been proposed to release elastic tensions in discrete so-called glitches of (radio-pulse periods emitted by) pulsars. Can the crust of a neutron star -- of thickness several Km -- store large enough tensions, corresponding to decreases in the moment of inertia by $10^{-7\pm2}$ when released? How similar is their behaviour to that of the crust of Earth? When Eckhard Krotscheck, Hans Heintzmann, and I tried to answer these questions, 
we noticed that such crusts are not expected to behave like steel -- even though their shear modulus is of comparable order -- but more like jelly, because their compressibility is some $10^2$ times less (than for steel). We then noticed that the crust 
should behave (to first order) like an incompressible (electron) fluid neutralized by a fragile lattice of positive ions, and that the standard treatment in the literature should not apply. The fundamental equations change from those of a solid body to 
those in the `figure' (below). The crust is expected to quake or tear internally, not starting at its edges, and to possibly give rise to frequency noise of the pulse-arrival times, (not to glitches). Our MS, proudly submitted to the Ann. of Physics in April 
1975, has never been criticised, and was finally declared `lost', after two years. We were at that time involved in different projects, and gave up. (Glitches happen with unmeasurably small heating of the neutron star. They are therefore explained by sudden 
couplings of the faster spinning neutron superfluid to the star, not by its stepwise shrinking: see my 1998a review.

\newpage

{\bf 2. A Black Hole's Entropy} S has been defined by Steven Hawking in 1974,
 in analogy to the irreversible thermodynamics of boxes;
his expression predicts a huge increase of S during  formation, much larger than the entropy of the whole Galaxy for a stellar-mass black hole. Instead, the entropy of a (hypothetical) shrinking star is known to decrease monotonically as it becomes a white 
dwarf, or neutron star: the necessary increase in total entropy escapes in the form of radiation. Hawking's expression equals the entropy of the hole's randomized evaporation radiation, as though evaporation was an isentropic process. For these reasons,
 I have suggested considering entropy a fourth `hair' (besides mass, spin, and charge of a black hole) which measures its age, and which equals Hawking's expression asymptotically on the evaporation timescale. Apparently, my 1976b `Letter to Nature' failed 
to compete with Hawking's fame (e.g. Moss, 1996), perhaps assisted by a dozen of serious uncorrected typos in the printed version; and so did 1976c.
\medskip

{\bf 3. The planet Venus} was known (in 1977) to have its spin phase-locked to Earth to within four significant figures, at a spin period of 243.1d. I.e. at every near encounter, Earth sees the same hemisphere of Venus (when looking through its cloud layer, at suitable frequencies). Earlier authors had explained this effect by dissipative tidal coupling 
(Gold \& Soter, 1969). At the same time, they had found the solar atmospheric torque capable of propelling the Venusian atmosphere, which superrotates quasi-rigidly at a significant 
fraction of its sound speed.
 When I re-estimated all torques that had been exerted on the planet throughout its history, I found it plausible that Venus has been born prograde like (almost?) all the other bodies in the solar system, with a spin period of order 5h, then spun down tidally 
by the Sun on the timescale of the formation of its atmosphere, some $10^9$yr, and subsequently spun up in the retrograde sense by above-mentioned (thermally induced) atmospheric torque. If so, Venus has already traversed 10 deeper spin resonances with Earth in the past, and is unlikely to be caught in the $11^{th}$ resonance. My prediction was confirmed by Shapiro et al in 1979 who published the improved spin period of  $(243.01\pm0.03)d$ (which differs from the resonance period of 243.16d).

\vspace {4.5 cm}

{\bf 4. The Speed of a Signal} --  according to Einstein's theory -- should never exceed the speed of light. Yet there are occasional reports, even in leading journals (like `Nature'), that under certain conditions, signals can do so; e.g. via the Casimir 
effect, or via tunneling. (Replies to such reports have not been printed). A nice survey of seemingly superluminal phenomena can be found in (Chiao, 1996). In 1978, Eckhard Krotscheck and I wondered whether superluminal signal speeds were permitted by 
relativistic equations of state, in particular by those applying to neutron-star matter, and found that they could not be ruled out from first principles, but could be ruled out when certain additional (reasonable) conditions were satisfied. Important was the 
insight that the speed of a signal is different from both its phase velocity $\omega/k$ and group velocity $d\omega/dk$ (where $k$ = wavenumber, $\omega$ = angular frequency), rather given by its `front velocity', viz. the infinite-wavenumber limit of its phase velocity (because it 
is the shortest wavelengths that count when a wave packet is to be cut off).

\medskip

{\bf 5. Phase Diagrams} of matter describe its state (solid, liquid, gaseous molecular, gaseous atomic, plasma) for varying pressure p and temperature T, say. In 1983, Marko Robnik and I were interested in the degree of ionization of hydrogen at high T and 
p (considered of relevance to the moving emission lines from SS 433), and minimized a suitable expression for the Gibbs free enthalpy to extend the known part of the phase diagram. In this process, Robnik found a (second) critical point, at (p,T) = ($10
^{5.38}bar, 10^{4.28}K$), as well as a phase transition (from vapour to liquid metallic) just below it (in temperature). In subsequent years, other groups confirmed our findings with more high-brow methods. When this high-T phase transition was first 
explored in the lab, in 1996, its discoverers Weir et al spoke of a `surprise'.

\vspace{5 cm}

{\bf 6. Astrophysical Jets} are observed as elongated emission features, often predominantly non-thermal, with (lobe) elongations of order 5:1, knotty beams of typical opening angle $10^{-2}$, variable, broadband core spectra (from radio to hard $\gamma$-
rays), core/lobe power ratios of order $10^{2}$, and often superluminal expansion rates (of their emission knots). They are produced by four classes of objects: (i) active galactic nuclei, (ii) newly forming (pre-T-Tauri) stars, (iii) young binary neutron 
stars, and (iv) young binary white dwarfs (inside planetary nebulae); cf. figure above. The beams are thought to be supersonic; but little consensus has developed over the years about whether they (all) consist of `bullets', i.e. are `hard', heavier than 
their surroundings --  like the water droplets in a (swinging) lawn sprinkler -- or whether they are `soft', lighter than their surroundings -- like the hot air from a hair drier. In the first case, the shape of the beam maps the history of firing directions 
whereas in the second case, it maps its continual interactions with the (heavier) windy ambient material, and allows for multiple refocussing. My own distinct preference has been in favour of soft beams, ever since my Letter to Nature with Gopal-Krishna 
in 1980, cf. our 1981 version as well as my communications in 1984a, 1987b, 1989a, and 1996b, whereas the scientific community has largely followed the ideas collected in Begelman et al (1984). A recent convergence may be indicated in Begelman et al 
(1994).

\medskip

{\bf 7. The Beam (bulk) Velocity} in the astrophysical jet sources -- introduced under alternative 6 -- has been particularly controversial. Guided by the impossibility of in-situ (post-) acceleration of the radiating electrons (and positrons?), my own 
preference for the jet substance has always been extremely relativistic pair plasma, of typical Lorentz factor $\gamma = 10^{4 \pm 2}$, generated by the central engine (in localized magnetic reconnections). It seems equally required by the extremely hard AGN 
spectra (reaching and exceeding TeV photon energies, to be understood as inverse-Compton losses of the forming-jet substance), and by the superluminal knot velocities, as by the high-frequency (UV) synchrotron turnovers in the knots and heads of the jets. 
Jet channels are rammed by quasi-weightless, inmiscible (magnetized) pair plasma: 1996b. In contrast, the scientific community has been reluctant, for some 15 years, to consider jet bulk Lorentz factors larger than some ten for the extragalactic radio 
sources, cf. Begelman et al (1984, 1994), and much less so for the jets from young stars -- despite the barring momentum and multiple refocusing problems resumed in alternative 12. Numerical simulations run into all sorts of instabilities unless they are 
designed in an axi-symmetric way with a controlling magnetic flux of non-reversing sign.

\vspace{5 cm}

{\bf 8. Astrophysical Jets} might consist of various substances, preferentially of hydrogen (which prevails in the interstellar medium), or (relativistic) pair plasma. A corner stone has been the Galactic neutron-star binary SS 433 -- alternatives 19, 20 
below -- which has served as a prototype jet source whose beams were believed to emit hydrogen and helium recombination lines, and X-ray lines from (hydrogen-like) iron, nickel, magnesium, calcium, silicon, sulfur, argon, and neon, i.e. to consist of local 
galactic matter. Instead, I have been convinced -- as stated above and in 1996b, 1998a -- that (in particular) the multiple (re-) focussing and never splitting of jets requires a different, quasi-weightless, inmiscible substance, and that all jet sources are capable of 
generating it: pair plasma.

\medskip

{\bf 9. The Beaming Pattern} in the astrophysical jet sources has often served as a diagnostic tool, to estimate the bulk Lorentz factor, under the assumption that the charges were radiating isotropically in their comoving frame. This assumption can be very 
unrealistic, as a power-law energy distribution of charges (with cutoffs) deviates grossly from power law in its center-of-mass frame: a frame with (relativistic) isotropic particle velocities need not exist. Rather, the beaming pattern of (accelerated) extremely 
relativistic charges reflects the integrated distribution of their tangents. When the electrons are diverted, by curved channel walls and/or (small-filling-factor) `obstacles' intruded into the channel, they radiate into a wide cone of forward directions, 
possibly as wide as $90^o$ in opening angle, but hardly in backward directions, independently of the size of their (large) Lorentz factor. This simple interpretation has been widely ignored even though it was rediscovered by Lind \& Blandford in 1984.

\vspace{6.5 cm}

{\bf 10. The bright Knots} in the lobes of the astrophysical jet sources -- discussed already under 6. to 9. -- have found various interpretations in the literature. How high are their pressures, what do they consist of, how are they formed? In my understanding, 
they are the sites of deceleration, not acceleration, of a certain subset of charges, the places where the -- otherwise collision- and loss-free -- jet substance encounters flow resistance on the environment. Often the (small subset of) radiating 
charges in a (non-terminal) knot have short lifetimes. They lose a large fraction of their energy by synchrotron radiation, but are re-accelerated and dragged along by their unperturbed neighbours via their frozen-in magnetic and electric fields so that 
behind the knot, the same charges continue, sharing the non-radiated fraction of the energy with which they had entered the knot. The situation can be likened to a company of soldiers in war who continue marching through centers of resistance: the share of 
the wounded comrades is taken by the rest, the total energy (of marching) is each time degraded, but often only weakly so.

\medskip

{\bf 11. In-Situ Acceleration} of charges in the jets (as well as in supernova remnants: Reynolds \& Chevalier, 1984) has been invoked by most of my colleagues in numerous publications, based on seminal ideas of Enrico Fermi and supported by a near-uniform 
consensus of the world's best theorists. I don't think, though, that it is anywhere required, nor do I think that it is consistent with the second law of thermodynamics. To be more specific: many of the calculations which have been done in support of in-situ
(second-order Fermi) acceleration have been performed in the test-particle limit (of negligible inertia of the charges), where they tend to be idealised (e.g. by ignoring magnetic fields) but correct. Whenever high efficiencies (in excess of 1\%) are 
required, however, the test-particle regime is no longer warranted, particle scatterings pick up inelastic contributions and recoil degradations, losses take place (of both high-energy particles and scattering waves) from the acceleration region, and an 
approach of thermal equilibrium sets in: 1984b, 1989b, 1990d, 1993a, 1996b, and Falle (1990). The deep reason against the feasibility of such a preferred redistribution of particle energies is, I think, the Second Law which becomes applicable as soon as 
(more than two) particles interact with each other many times without an external control of the scattering phases (such as in terrestrial accelerators).

\vspace{5.5 cm}

{\bf 12. Bipolar Flow} has originally been the name for an outflow scenario from a star-forming region, but is nowadays also used for the other astrophysical jet sources. When bipolar flows in the original sense were first discovered, they were not expected 
to follow a similar working pattern to the -- much larger and much more energetic -- extragalactic radio sources, because apart from the size difference, the latter emitted mainly non-thermal radiation, and were believed to be powered by black holes whereas 
the former are powered by forming stars. Models were therefore constructed that allowed a forming star to focus its wind into two antipodal lobes, see above figure. Such models have failed to convince me, for various reasons spelled out by Hajo Blome 
and myself  in 1988, also in 1984a,1987b: On the one hand, the putative black hole in the center of an active galaxy is most likely its burning central disk: 1996a, i.e. is not too different from a forming star; and on the other hand, the densities surrounding 
a pre-T-Tauri star tend to be some 10 orders of magnitude higher than extragalactic densities, implying that most of the outflow energy from a young star is thermalized, and its non-thermal emisson is largely drowned. It should, therefore, not take by 
surprise that the two classes of sources have in general quite different spectra. But there do exist stellar radio triple sources which look indistinguishable from their big extragalactic brothers, also one-sided core radio jets like in HH 111: Reipurth 
\& Heathcote (1993). The lobes are likely to show up at very low radio frequencies. And in any case, I see no way of multiply focusing stellar winds into outflow channels of opening angle $\ga 10^{-2}$, or even narrower. Bipolar flows are miniature copies of the extragalactic jet sources: 1996f.

\medskip

{\bf 13. 3C 273} is one of the nearest and best-studied (strong) extragalactic jet sources, probably one of the youngest as well. It shows only one jet and lobe -- at a dynamic range reaching $10^{3.7}:1$ -- interpreted to be approaching us at a small angle 
to the line of sight, cf. Bahcall et al (1995). Among the over $10^2$ well-mapped (extended but compact) extragalactic radio sources, there are only $\ga$2 further strictly one-sided sources. Gopal-Krishna and I (1986) interpret this sidedness as due 
to the finite relativistic light-travel time -- which records the counter jet at a younger stage -- combined with a low-density galactic (cosmic-ray) halo (through which the jets ram their hardly visible channels) and an extremely relativistic bulk speed of the jets' (pair) plasma.

\newpage

{\bf 14. The Motion of the Beam Particles} in the jet sources tends to be simulated numerically by hydrodynamic, or magnetohydrodynamic calculations. In reality, this motion may be cold and ordered, guided by frozen-in transverse magnetic (toroidal) and 
transverse electric (radial, Hall) fields. Such a field configuration -- an ordered $\vec E \times \vec B$-drift -- may develop naturally (and stably) due to the jet's formation in crossing a deLaval nozzle (Blandford \& Rees, 1974) and a (cooling) dense photon bath followed by charge-asymmetric
wall friction (which generates an axial current density, hence toroidal magnetic field) combined with the Hall effect (which redistributes the axial separations of opposite charges infinitesimally, to generate an electric field of equal magnitude to B): 
1989a, 1996b. Only such a field-guided flow has the properties of being both loss-free (in straight channel segments, ignoring inverse-Compton losses on the ambient photon field) and coherent (so that a forming jet never splits).

\vspace{1 cm}

{\bf 15. The Central Engine of an Active Galactic Nucleus} is commonly believed to be a supermassive Black Hole, after Donald Lynden-Bell (1969) and Martin Rees (1977). Alternative, more conservative interpretations -- like star clusters, or a supermassive 
star (spinar, magnetoid) --  have been rejected by (i) variability arguments: a huge power is modulated on short timescales, (ii) an absence of (spin) periodicities, (iii) energy arguments: is there enough (nuclear) fuel?, and (iv) stability arguments:
any other type of engine should evolve into a black hole. 
Yet (i) the time-integrated QSO activity should have left more than $10^4$times the mass in the centers of galaxies in the form of black holes than there is now in our cosmic neighbourhood, and (ii) the 
broad-line region signals as much outflow of mass on average as spirals in, both inconsistent with a monotonic accumulation of mass in galactic centers. Further difficulties of the Black-Hole model are (iii) the hard spectra -- occasionally peaking above TeV photon-energies -- , (iv) the high pair-plasma opacity -- which would forbid jet-formation -- , and (v) the inverted QSO evolution -- having the brightest sources at the beginning. 
Besides, the last word is not yet spoken about how much of the cosmic helium has been formed in the first three minutes (after the bang), before the quasar era. For these reasons, I favour a less massive variant of the spinar model: the (nuclear-) burning center of the galactic disk -- of mass several $10^6 M_{\odot}$ and solar-system extent -- which is continually refuelled, and 
evolves (radially inward) through main-sequence hydrogen burning towards explosive He and `metal' detonating all the way to iron, thereby ejecting the ashes of nuclear burning: 1979a, 1987b, 1996a,b. Such a `Supermassive Magnetized Core', or `Burning Disk' is expected to look highly variable 
-- its power being dominated, at each instant, by only few large detonations -- non-periodic, and self-limiting (like big oil fires quenched by the injection of explosives). Moreover, Turnshek (1988) has stressed that the ejecta of QSOs are strongly ($10^
2$-fold) iron enriched (compared to solar abundances), like the ashes of explosive nuclear burning.

\vspace{7.2 cm}

{\bf 16. The Broad-Line Region of a QSO} (=Quasi Stellar Object) is the region around an AGN which emits spectral lines whose broadening signals huge velocities, of up to one tenth the speed of light. As stated under 18., my preferred interpretation of these 
huge velocities is ejection. In any case, there is the problem of how the high-pressure line-emitting cloudlets -- of tiny volume-filling factor -- manage to cross the BLR (supersonically) without exploding; which they would if the latter consisted of 
interstellar matter at temperatures in the vicinity of $10^8K$, as assumed by Krolik et al (1981). My proposed solution of this apparent paradox is that the BLR is not filled with matter of T = $10^8K$ but rather with matter of T $\gg 10^8 K$: relativistic 
pair plasma, generated in localized coronal discharges and escaping into the two antipodal jets; so that the cloudlets move subsonically, confined by the pair-plasma's quasi-static relativistic pressure, like in the Crab Nebula: 1979a, 1987b, 1996b.

\medskip

{\bf 17. The Big Blue Bump} (= hard UV part) of a QSO spectrum is often the peak of its emitted spectral power (= power per logarithmic frequency), and is thought to be the (largely thermal) main output of the central engine. [At times we receive even more 
power at and beyond TeV photon energies, probably due to (huge) inverse-Compton losses of the escaping electron-positron pairs; but we do not know their emission `cone', i.e. the fraction of the sky into which this very hard radiation is beamed: its total 
power may be comparable]. Where does the BBB come from? As a naked black hole does not radiate (except for its completely negligible evaporation radiation), the BH model is forced to attribute the BBB to the surrounding accretion disk. The `Burning Disk' (=BD) model, 
instead, attributes the BBB to nuclear detonations -- plus in-situ burning of pairs -- of the innermost disk: 1987b, 1996b. Note that unlike the BH model, the BD model explains the (90\%) radio-quiet QSOs as those in which the  relativistic electron-positron pairs -- generated in coronal discharges (powered by magnetic reconnections) -- do not manage to escape into cooler regions before being burnt (by collision losses on the central BBB).

\vspace{6.5 cm}

{\bf 18. The Motion in the Broad-Line Region} could be rotation, outflow, or infall, as long as the only observational knowledge is velocity broadening.
If it is rotation, the implied enclosed mass which binds the motion would often have to be much larger than that implied by radiation at the Eddington limit. If it is infall, there would have to be some mechanism at work that pulls matter radially inward, 
with zero net angular momentum, often rapidly followed or preceded by outflow, at velocities reaching 10\% of the speed of light -- a highly unlikely occurrence. Very similar spectra have been obtained from the (young, variable) YY Orionis stars where 
occasionally, blueshifted and redshifted self-absorption take their turns even in different parts of the same spectral-line multiplet, naively looking like a fountain, not like a stellar wind.
Returning to the first case (of an AGN), infall has alternatively been inferred from the delays with which different spectral features arrive during variable epochs. On the other hand, both occasional absorption events and maser emissions speak in favour of outflow. 
In 1988b, I have argued that all these phenomena can be interpreted purely as outflow by taking proper account of radiation transfer: Line self-absorption can happen selectively in the approaching (blue-shifted) gas column, and so can the delay from multiple 
scatterings off small-filling-factor filaments. (A technical error in 1988b is that I assumed the central galactic disk to be transparent: we should only be able to see the front hemisphere). Outflowing wind zones can account for all the observed (spectral 
and time-of-arrival) observations.

\smallskip

{\bf 19. The Moving Emission Lines of SS 433} -- the $433^{rd}$ star in the catalogue of variable stars by Stephenson and Sanduleak (1977) -- look almost like incoherent recombination lines from hydrogen and helium (in their frequency and intensity ratios), 
periodically both blue- and redshifted corresponding to a significant fraction of the speed of light. 
Mordecai Milgrom interpreted them in 1979 as emitted by matter (`bullets') moving along the presessing generators of a circular cone, at a quarter of the speed of light. The system SS 433 is a key object for our astrophysical understanding because it combines 
the (rare) presence of a non-accreting (young, massive) binary neutron-star at the center of the $10^{4}$-year old supernova remnant W 50 with a Galactic jet source and an emitter of a broadband continuum plus a multitude of emission lines, both `stationary' 
and moving, at IR, optical, and X-ray frequencies (from [hydrogenic] iron, nickel, magnesium, calcium, silicon, sulfur, argon, and neon). The emissions are linearly polarized, and show several periods: orbital, precessional, nodding, and further 
beat frequencies, as well as correlations on all timescales down to minutes. 
Nevertheless, almost all derived properties of SS 433 are controversial, such as its distance (by a factor of 2), energetics (by a factor of $10^3$), orientations, and jet composition: 1979c, 1981b, 1985b, 1987d, 1991a, 1996b, 1998a; also Jonathan Katz's oral summary 
of the Washington-University conference at St. Louis during the $13^{th}$ Texas Symposium at Chicago (1986). What most people believe to be emitted by the jet substance may in reality be emitted by channel-wall material which is dragged along by the escaping pair plasma.

\vspace*{6.5 cm}

{\bf 20. The Jets of SS 433}, mapped at radio frequencies on all length scales between $\la 10^{14}cm(d/3Kpc)$ and $10^{20.2}cm(d/3Kpc)$ and at X-rays on the large scales only, have a similar appearance to all the other jet sources discussed in alternatives 6 to 14. 
Yet as already mentioned in 19., they have been widely interpreted as consisting of `bullets' made of interstellar matter rather than of extremely relativistic pair plasma, due to their (mis-) identification with the line-emitting source -- which may instead be the multiply excited channel wall. A seemingly close correspondence with the evolving radio jets mapped on the VLA and VLBI scale had to assume a large distance (by a factor of 2), an opposite orientation and precession sense (to the motion of the optical filaments in W 50, and to the linear polarization of 
the optical continuum), and (mis-) identified the peak of the radio emission with the central source, assuming it to be the strongest emission `knot' at all times. 
In 1987d, I argued against this interpretatation, and converted it subsequently into a 100 \$ bet; which has not been responded to. In the meantime, the Galactic superluminals (GRS 1915+105 and GRO J1655-40) hold an answer.

\vspace{0.6 cm}

{\bf 21. The $Ly \alpha$ Forest} (of QSO absorption lines) are the over a hundred narrow, blueshifted $Ly \alpha$ absorption lines which can be identified in every QSO spectrum. The intervening neutral hydrogen which gives rise to them must occur in `warm' 
($\ga 10^4$ K), low-column density (N(HI) $\ga 10^{14}cm^{-2}$), low-metallicity (Z $\la 10^{-3}Z_{\odot}$)  `cloudlets', or filaments of narrow velocity dispersions ($\Delta v \la 10 Km/s$) whose spatial distribution is similar to that of galaxies but 
with a considerably larger surface filling factor, corresponding to oversized galactic haloes. If confined staticly by the (hot) intergalactic medium, the `cloudlets' would be expected to have long since evaporated. For these and other reasons, Marita Krause 
and I interpreted them in 1985 as (slightly) supersonic filaments ejected by the nuclei of galaxies during their active epochs -- their ejection being evidenced by the BLR (alternative 18); see also my report in 1987c, agreed with by Wallace Sargent in 
his Chicago Texas-Symposium workshop-summary talk.

\vspace*{6.5 cm}

{\bf 22. Galaxies in Clusters} can collide, like stars in a star cluster, or like particles in a container. How different is the evolution of galaxies inside dense clusters from those outside of such, due to encounters? During near encounters, are galaxies 
stripped of their gas, or do they accrete, or merge? Galactic cannibalism has been invoked by Jerry Ostriker for the formation of giant cD galaxies, at the centers of large clusters, recognizable by their massive haloes ($M \la 10^{14} M_{\odot}$), and 
multiple nuclei (several of which have turned out to be chance projections). When one calculates the expected collision time of galaxies from the statistical-mechanics formula $\tau = 1/n \sigma v$, and tests it on the cD galaxies (with their few mergers 
during the age of the Universe), one predicts a merging probability $\tau / t(universe)$ smaller than $10^{-3}$ for typical cluster galaxies, i.e. a rather rare occurrence: 1987c. The systems of (partial, antipodal) emission shells seen in particular around 
isolated elliptical and S0 galaxies, out to distances of several hundred Kpc from their center, must then be explained in a different way, e.g. by nuclear ejection: alternative 21. From the Heidelberg conference `Dynamics and Interactions of Galaxies', 
I was sent home during the first coffee break, on 29 May 1989, by my former friend Alar Toomre, because I was `not invited'. (The meeting had been designed to take place without C.C. Lin and myself). In 1996, Moore et al have proposed that galaxy `merging' 
should be replaced by galaxy `harassing'.

\bigskip

{\bf 23. The Wisps in the Crab} are some $\ga 9$ thin, bright optical streaks seen almost symmetrically around the central pulsar (PSR) -- though much fainter on the `receding' side -- transverse to the long axis of the `nebula', and erratically moving 
at a significant fraction of the speed of light; cf. Hester et al, 1995. In projection, they are emitted so near to the PSR that in 1977b, I could not see a mechanism that would produce them at that location. Instead, they looked to me like a (coherent 
synchrotron) phenomenon taking place at the inner edge of the nebula, where the radially $\vec E \times \vec B$-drifting electron-positron pairs from the PSR join the escaping queue of the nebula, in transit from straight-line motion to gyration, via eel-like
meandering. They would thus move similarly to the relativistic electrons in the 1977 Stanford free-electron laser, and preferentially radiate in preferred near-radial directions, appearing like transverse streaks to a distant observer.

\newpage

{\bf 24. The Crab-Pulsar Wind} is the medium blown out by the PSR, and slowing its rotation, much more so than does its electromagnetic radiation. Most likely, the wind is extremely relativistic, consisting of (i) (strong) low-frequency (30 Hz) waves, (ii) 
extremely relativistic electrons and positrons (of Lorentz factors between $10^5$ and $10^9$, sharing the initial energy of the waves), and (iii) some residual magnetic flux from the rotating dipole at the center, of opposite sign in the two hemispheres. The wind reveals 
its existence indirectly, by the emitted radiation of the outgoing and stored charges in the nebula -- both synchrotron, and synchro-Compton -- and more directly by having post-accelerated the thermal filaments ejected by the supernova in the year 1054, 
by some 8\%. This post-acceleration is over 30 times stronger than the wind's radial momentum, motivating Eckhard Krotscheck \& me (1980) to update the master paper on the Crab by Rees \& Gunn (1974), trying to find a consistent description of this well-studied,
multiply overdetermined astrophysical source. The answer to this post-acceleration problem at which we arrived during our more than biennial study -- a `bath' of strong 30 Hz waves reflected over 30 times from the thermal filaments -- differed from 
earlier treatments in quite a number of details, and has fed my interest in PSRs, SNe, SNRs, PSR nebulae, and BLRs ever since: 1985c, 1990a, 1996c, 1998a.

\vspace{1 cm}

{\bf 25. The Cosmic Rays} are relativistic ions and electrons flooding our cosmic neighbourhood, with a chemical composition similar to that of local galactic matter (except for a tenfold underabundance of both hydrogen and helium, an enrichment in [rare] spallation products, and a slight preference for elements 
with a first ionization potential below 9 eV), with particle kinetic energies from below rest energy all the way up to $10^{20.5}eV$, distributed as a broken power law which peaks (in spectral power $E^2 \dot N_E$) at an ion Lorentz factor of 5.  The 
location of the corresponding peak for electrons is below 0.2 GeV, with a controversial flux. Where do the cosmic rays come from? As the charges are forced on curved (gyration) orbits by the ubiquitous (solar-system, Galactic, and/or cluster) magnetic fields, 
their arrival directions contain little information about the location of their sources in the sky. Even worse: at the highest energies, when the arrival directions should not differ much from the source location, their distribution is almost isotropic. 
Most workers in the field believe that the cosmic rays can reach their fantastic energies in multiple collisions with certain magnetized cosmic plasma clouds, via shock acceleration: alternative 11. Instead, if (efficient) shock acceleration violates 
the second law, the only known environments capable of reaching the highest particle energies (in single-step boosts) are (strongly magnetized, fast-spinning) neutron stars and their accretion disks: 1976d, improved in collaboration with Nigel Holloway \& 
Yi-Ming Wang: 1978, and extended in 1984b, 1989c, 1990d, 1992c, 1993a, and 1998a to include slingshot-like radial ejection from the inner magnetosphere, after having been `strained' by impacts of a filamentary `drizzle' from the returning tail of SN ejecta which have failed to reach escape velocity.

\vspace{7.0 cm}

{\bf 26. The Highest-Energy Cosmic Rays} -- with particle energies E above $10^{19}eV$ -- cause particular headaches to theorists because of the high voltage required for their acceleration, and because of their short lifetimes, due to radiative losses. Often 
they are only considered of diagnostic importance, because of their comparatively low flux density near Earth ($10^{-7}$). But contrary to the low-energy cosmic rays which are stored in the galactic-disk magnetic fields for $10^{7 \pm 0.5} yr$, the UHE 
cosmic rays traverse the galactic disk almost in straight-line motion, i.e. are `stored' some $10^{4.5}$ times shorter, implying that their sources must invest almost 1\% of their power into them (if Galactic), a significant fraction. On the other hand, their (almost) 
isotropic arrival directions seem to signal an extragalactic origin: Hillas (1984). But then the putative cosmic-ray boosters would have to refill all of space, not just the disks of early-type galaxies, with an even larger share at UHE energies (due to 
enhanced collisional losses on the cosmic background radiation at higher energies). Isotropy can also be achieved by many nearby neutron stars, in particular by those with spin periods in the (short) msec range: 1993a, 1998a.

\vspace{0.7 cm}

{\bf 27. The stellar-mass Black-Hole-Candidates} (BHCs) are X-ray binaries containing a compact component whose mass -- judged from the line-of-sight velocity oscillations of the visible companion -- is of order $(6 \pm 2) M_{\odot}$, larger than the stable 
mass limit of a neutron star: White \& Marshall (1984), Lewin et al (1995). On the other hand, their X-ray light curves mostly range between the Eddington luminosity of a neutron star (of $1.4 M_{\odot}: 10^{38.5}erg/s$) and almost undetectability ($\ga 10^{31}$
erg/s), unlike a black hole surrounded by a standard-type accretion disk. Their erratic optical light curves, and large-equivalent-width emission lines signal an extended, luminous windzone blown by both components. During (soft X-ray) `low' state -- nowadays called `hard' state -- the 
spectra can have their energetic peaks above 1 MeV (!), reminiscent of cooling sparks. In all other properties, the BHCs appear indistinguishable from neutron-star binaries (as a class), such as: (i) high-low state X-ray variability, (ii) X-ray dipping, (iii) third (precessional) period, (iv) type II X-ray bursts, (v) shape 
of X-ray-noise power spectra, (vi) polarized optical emission, (vii) superhump-type excursions of the orbital period at outburst (alternative 71), (viii) Li absorption, (ix) formation of (radio) jets, (x) having supersoft spectra, and (xi) being super-Eddington. For these and other reasons, I found it more plausible, 
in 1979b, that the BHCs involve neutron stars surrounded by massive, self-gravitating accretion disks, received via mass transfer from their (formerly) massive companion; see also Daniel Fischer and myself: 1989. Such massive disks cannot occur around white dwarfs, because of their much higher accretable Eddington mass rate (of $10^{-5} M_{\odot}/yr$). The disks rotate rigidly for most of the non-transfer time, hence give rise to the long intervals of quiescence (decades) in the transient low-mass X-ray 
binaries; 1996c, 1996d. They permit super-Eddington emissions (by providing high-density fuel) as encountered in at least 3 BHCs, and are observed as `SuperSoft (X-ray) Sources' (SSS) in the process of formation, alternative 72. Massive disks have also been considered by Krolik (1984).

\vspace{5.5 cm}

{\bf 28. Line-Of-Sight (LOS) Velocity Oscillations} of emission and/or absorption lines are routinely used to determine the `mass function' of stellar binaries. This method is among the most reliable ones for mass determination in astrophysics, being based on 
Kepler's third law, but gets unreliable as soon as the -- partially corotating -- windzone of the two stars gets opaque in the concerned line. Observationally, this situation can occur for mass-loss rates in excess of $10^{-10}M_{\odot}/yr$, and is signalled 
by a noisiness of the LOS-velocity curves. In 1990, Indulekha, Shylaja \& I (finally) managed to get a paper published which discusses a few well-studied binaries for which unreliable data had been obtained in the literature, among them Cyg X-1 and SS 433. At the same time, Wolf-Rayet stellar winds are found to be centrifugally driven (rather than radiatively -- an interpretation which is at variance with the radial-momentum balance --), and non-negligibly also the solar wind.

\vspace{0.7 cm}

{\bf 29. Double-peaked Emission Lines} from compact binaries -- or QSOs -- are often interpreted as due to Keplerian motion, and are thus used for a mass determination of what is contained inside of the emission ring. Problems with this interpretation are 
that the involved disks tend to be optically thick, i.e. emit blackbody radiation except for the possible presence of a hot(ter) corona, and that they are small. Even in the presence of a line-emitting corona, the (large) equivalent width of an emission line requires a minimum emission 
area which must not be larger than the projected ring area inside of which the disk's rotation speed amounts to the necessary Doppler-shifted share. In my 1989 communication with Daniel Fischer and in 1996b, I have used such estimates to argue that the 
split (broad) lines stem from an extended wind zone -- in both cases. No emission lines from disks are known to me.

\vspace{5.5 cm}

{\bf 30. Neutron-Star Dipole Moments} are estimated from the spindown of pulsars, interpreted as due to the emission of strong magnetic waves by an oblique rotator: 1986, also: Kulsrud et al (1972). In a few cases, neutron-star surface magnetic-field strengths 
have been inferred from cyclotron emission, absorption, or scattering (at X-ray energies -- first recognized by Joachim Tr\"umper -- yielding values above $10^{12}G$). All other field-strength estimates of neutron stars are more indirect, by judging 
the strength of the torque acting between the star and its disk during mass accretion, or the fraction of accreted matter that is funnelled onto its polar caps. There has been considerable uncertainty, throughout the 80s of this century, whether or not the dipole moments of pulsars decay, grow, or are conserved, in particular in view of the question of why pulsars turn off at an age of $\la 10^{6.5}yr$. My own preference has been in favour of dipole non-decay: 1981a, 1988a. More recently, though, the weak 
dipole moments of the msec pulsars and the different fine structures of different pulsar glitches make me believe that pulsars are born with strong higher multipole moments (obtained during the supernova explosion, via a toroidal bandage: alternative 52), which decay during 
their lifetimes, thereby modulating (and even enhancing) the dipole: 1994, 1998a.

\vspace{0.7 cm}

{\bf 31. Pulsar Winds} are thought to consist of electron-positron pair plasma, in particular after Malvin Ruderman and collaborators (cf. 1975), the pair plasma receiving most of its power (just) outside the corotating magnetosphere, near the speed-of-light 
cylinder. This post-acceleration was initially thought to happen according to the Gunn-Ostriker mechanism (of boosting via the strong, forming magnetic dipole wave, cf. Kulsrud et al, 1972), but came into disrepute after various instability claims (by 
Wilhelm Kegel, Estelle Ass\'eo, and others) which indicated that the coexisting plasma would quench the wave (if superluminal). Such claims could not convince me, however, because a medium that absorbed the strong wave would thereby gain its radial four-momentum, hence be 
post-accelerated: 1986, 1994.

\vspace{6.0 cm}

{\bf 32. Pair-Plasma Winds}, as discussed in the preceding alternative, should not only be formed by pulsars but rather by all fast-rotating, strongly magnetized (neutron) stars, as long as their magnetospheres are not quenched, e.g. by the wind of a very 
near companion. When the extremely-relativistic pair plasma cannot escape into all directions, a twin-exhaust scenario is expected to form which funnels it into two antipodal (supersonic) jets: 1986, 1996b, 1998a. More than a dozen such neutron-star driven jet sources 
are now known, among them SS 433.

\bigskip

{\bf 33. Neutron Stars form} from massive stars, massive enough for gravity to overcome the electrons' repulsive degeneracy pressure -- which can stabilize a (low-mass) star in the form of a white dwarf -- and to compress matter to more-or-less nuclear 
density. What progenitor stars end up as neutron stars?


\noindent{Estimates of the critical mass at birth of a star for its ending up, respectively, as a white dwarf or a neutron star, have changed quite a bit over the years, upwards from 3 to 10 $M_{\odot}$ and then 
back down to $\ga 5 M_{\odot}$, among others by a comparison with pulsar-, supernova-, and supernovaremnant-birthrates: 1985c, 1988c, 1998a.}

\vspace{5.5 cm}

{\bf 34. Neutron Stars are observed} as (radio) pulsars, variable X-ray sources, and jet sources. Whereas most of the X-ray emitting neutron stars are thought to accrete matter from a near companion -- namely all non-ejectors -- hence are members of stellar 
binary systems, the pulsars are practically isolated neutron stars, emitting their coherent radio pulses (deep inside the magnetosphere) because of their strong, bunched forming winds. The pulsars have large peculiar velocities, at least $\la 10^{2.5}$
Km/s -- recently claimed to reach, or even exceed $10^3Km/s$, partially based on somewhat uncertain dispersion-measure distances (which I mistrust: alternative 46) -- whose origin is not well known. Do they receive large kick velocities during the supernova 
explosion that gave birth to them? In any case, several of the youngest pulsars (like the Crab) are born far above (their birthsite in) the Galactic disk, at $\la 200 pc$, a height to which they got most likely via a `runaway' (stellar) system which 
received its kick during the supernova explosion of the first (initially most massive) star. I therefore tend to think that essentially all neutron stars stem from (massive) multiple-star systems: 1985c, 1998a.

\medskip

{\bf 35. Pulsar Beams} can only be observed indirectly, via their downstream bow shocks, and statistically, by guessing their latitudinal pulse profile from the observed longitudinal one, and from theory. Initially, the simplest assumption seemed to be 
(circular) pencil beams, in accord with the pulses being emitted tangentially to those dipole fieldlines which come from the two polar caps, being visible from roughly 20\% of space. But then there would be more pulsars than supernovae, and there should exist 
supernova shells containing a pulsar nebula without a pulsar. Moreover, refractive-scintillation events have occasionally signalled large emission areas, comparable to a significant fraction of the speed-of-light cylinder; and it is not clear whether 
or not pulsars are strictly quiet between pulses (and interpulses), at the $10^{-4}$ intensity level. Their antenna patterns may be spiky, with a large dynamic range, and fan-shaped when coarse-grained such that we see most of the very near pulsars (within 
$\la 10^2 pc$) but lose them out of sight increasingly with increasing distance: 1985c, 1988a, 1998a.

\vspace{6 cm}

{\bf 36. Pulsar Radio Pulses} are nowadays thought to be emitted `half-way' (in logarithmic scale) between the polar caps and the speed-of-light cylinder; but that has not always been so. Initially, the narrow pulses were taken as indicators of radiation 
from somewhere above the polar caps, perhaps near the local plasma frequency, though emissions near the light cylinder -- either tangentially or radially -- as well as near the polar caps were likewise considered, cf. 1990a. The mere existence of spiky 
pulses does not tie down their emission height anywhere between the neutron star's surface and the inner edge of the surrounding pulsar nebula because both the electrons and the photons travel at essentially the (same) speed of light. What may be diagnostic 
is their (frequency-dependent) polarization as a function of pulse phase, both linear and circular, as well as their correlated narrow-band subpulse- and broadband micro-structure. On theoretical grounds, highly coherent radiation is expected from an 
emission height at which the guiding magnetic field has decreased sufficiently in strength ($\ga 10^7$ G) to allow for a transient excitation of coherent gyrations around it, necessary to raise the radiative resistance way above that of curvature radiation, 
thereby permitting instantaneous brightness temperatures of $\la 10^{30}K$: 1985c, 1998a. Such emission would be low-pitch-angle synchro-cyclotron radiation.

\medskip

{\bf 37. Accreting Binary X-ray Sources} are observed abundantly in the sky -- both soft (involving a white dwarf) and hard (involving a neutron star) -- as pulsators, bursters, and flickerers. How does the matter which accretes onto the compact star get 
there: free-falling directly from the donor star's windzone, or after having been transiently stored in an accretion disk, due to its excess angular momentum? The literature distinguishes between `wind-fed' and `disk-fed' sources, depending on the donator's 
size relative to the Roche lobe; but there does not seem to exist any further property of the two classes of sources that would allow a distinction. Pulsators from both classes show rapid random-walk excursions of their spin periods around their -- widely 
distributed -- average values, understandable as exchanges of angular momentum with their disk via magnetic torques, but difficult to understand in terms of randomly changing wind momenta: 1985c, 1998a.

\vspace{6 cm}

{\bf 38. A Common Envelope} is often thought to form transiently around a massive star and its not-too-far neutron-star companion, i.e. for orbital periods less than about a year, leading to neutron-star spinup (`recycling') and to the formation of a msec 
pulsar: Ed van den Heuvel \& Dipankar Bhattacharya (1991). No such system has so far been identified in the sky, perhaps because of its short lifetime; but recycling would last long, because of the Eddington throttle, see next alternative. In any case, it is not clear to me whether such a common envelope around a neutron star does indeed form, or whether the neutron star will force the gas of the companion's envelope more or less gently into its flat accretion disk: 1985c. Is $\eta$ Carinae a system in this stage?

\medskip

{\bf 39. The msec Pulsars} seem to form a separate class of pulsars, with (i) distinctly lower magnetic dipole moments (B$_{\perp} \la 10^9$G), hence (ii) slower spindown, (iii) weaker spindown `noise', (iv) smaller peculiar velocities, and (v) with a high probability 
of having a low-mass companion ($M \la 0.3 M_{\odot}$). Can they be understood simply as born faster (than the `normal' pulsars), because of a weaker coupling  to the envelope of the collapsing progenitor star, hence weaker spindown during the 
supernova explosion, tighter toroidal bandage, much stronger higher magnetic multipoles, smaller kick velocity, and less efficient mass ejection of the supernova, (hence the formation of a low-mass companion, from incomplete mass ejection)? I have always thought 
so: 1980, 1985c, 1998a. Yet van den Heuvel \& Taam (1984) launched another scenario, in which a slowly spinning binary neutron star is `recycled' by mass accretion from its near companion. The problems with this alternative evolution are that (i) a neutron 
star surrounded by an accretion disk will not only tend to spin up due to accretion, but also tend to spin down due to magnetic disk friction, and due to blowing a pulsar wind, and that (ii) no single binary X-ray source has been found which would spin up 
secularly, at the maximum possible rate (as assumed in the recycling scenario, which would otherwise last intolerably long). Rather, all the known pulsing X-ray sources have pulse periods which fluctuate around temporary equilibrium values.

\vspace{6 cm}

{\bf 40. Neutron-Star Accretion} is thought to happen in the form of ionic motion along magnetic field lines onto their polar caps, suggested observationally by the existence of pulsing X-ray sources, and theoretically by the strength of the fields (whenever 
the strength can be reliably estimated: alternative 30). But only a small fraction of all binary X-ray sources show pulsations; do all the others have weaker fields? An absence of pulsations would ask for surface fields weaker than some $10^8 G$ whereas 
there are non-pulsing sources whose surface fields are expected to be much stronger than $10^{11} G$ by their abilities to (i) flicker on subsecond timescales, (ii) generate jets, (iii) emit hard $\gamma$-rays, and/or to (iv) polarize their wind zones. For 
this and other reasons, Mehmet \"Ozel, Nihal Ercan \& I explored the possibility (finally printed in 1987) that matter at the inner edge of an accretion disk gets decomposed into large fragments, by the violent intrusion of magnetospheric flux tubes. These 
fragments get increasingly diamagnetic by being increasingly squeezed on approaching the central star -- to white-dwarf densities - and are chopped up and partially evaporated by tidal forces, by collisions, and by `magnetic spanking'. Their non-evaporated,
flattish leftovers, or `blades', accrete onto a (rotational) equatorial belt, and give rise to a non-pulsing, soft X-ray source (because of a landing area much larger than the polar caps, with axial symmetry). An absence of pulsations need not mean 
an absence of strong fields, rather an inefficiency of evaporation, (e.g. because of a cooler environment). In 1993, Hsiang-Kuang Chang and I invoked such blades for an explanation of the (mysterious) $\gamma$-ray bursts via spasmodically accreting, nearby 
neutron stars: alternative 67.

\medskip

{\bf 41. The Non-Pulsing Neutron Stars, and msec Pulsars} are often believed to have weak magnetic surface fields ($< 10^{10}G$), because strong surface fields ($> 10^9G$) would funnel the accreted matter onto their polar caps, and would make the msec 
pulsars spin down faster (than they do). But as argued in the preceding alternative, accretion may take place largely in the form of non-evaporated chunks, or `blades', reaching the neutron-star surface along an equatorial belt, even for very strong fields, 
and pulsar spindown is thought to sense (only) the transverse (magnetic) dipole moment, not the higher multipole moments, whereas pulsar windzones reveal high densities of pair plasma -- some $10^4$ times the maximal Goldreich-Julian flux -- whose formation 
requires strong fields (via the Erber mechanism): 1985c, 1994, 1998a.

\vspace{6.0 cm}

{\bf 42. X-ray QPOs} is the term coined for `quasi-periodic oscillations' of an X-ray lightcurve, i.e. for bumps in its power spectrum. They have been often found in X-ray binaries, instead of (expected) clear periods (which would show up as sharp spikes), 
even though they may owe their existence to one or more strict periodicities, such as the spin of the central rotator, and/or the Kepler period of an orbiting clump. The discovered QPO intensities tend to exceed 1\%, and can reach 50\% in exceptional 
cases; which means that a significant fraction of the total (accretion) power can be involved in their generation. For this and other reasons, proposed disk instabilities have not been able to convince me -- as they happen too far above the bottom of the 
star's potential well (and can, moreover, easily average out) -- in contrast to the screening by a plasma-loaden magnetosphere which performs torsional oscillations; cf. my collaborations with Mehmet \"Ozel \& Nihal Ercan: 1987, and with Daniel Fischer: 1989.

\medskip

{\bf 43. An optical Pulsar in SN 1987A} in the Large Magellanic Cloud had been reported with a period of 0.51 msec (Nature 338, 234, 1989) when my 1990a NATO book  was printed. 
From the little I knew, I would not believe that a neutron star could spin with a period shorter than one msec (because of centrifugal instability). I therefore entered the pulsar into my table 1 with twice the reported period, pretending that its interpulse 
had been confused with the main pulse. The detection was later retracted, leaving Vladimir Lipunov (for the same reason) and myself happy that we had not believed in the (most likely) impossible.

\medskip

{\bf 44. Pulsar Torque Noise and Glitches} cause measurable irregularities of the predicted pulse arrival times, limiting the accuracy of pulsar clocks at a level that decreases with decreasing time derivative of the period (as a power law), and drops 
below the present long-term stability of the best terrestrial clocks ($10^{-14.6}$) for one or more of the shortest-period pulsars. Here the `torque' component of the `noise' tends to be the dominating limitation, stronger in the long run than its `frequency' 
and `phase' component as well as the discontinuous period jumps called `glitches', of relative amplitude $10^{-7 \pm 2}$, which are moreover obvious and can thus be corrected for. Attempts to understand these pulsar irregularities tend to consider the star's 
moment of inertia as the only `noisy', i.e. fluctuating quantity, cf. Michel (1991); whereas it is my impression that a fluctuating magnetic-dipole moment can occasionally contribute. Such fluctuations may be due to the secular decay of higher magnetic 
multipole moments acquired during the supernova explosion, by a pinching of the dipole via a toroidal bandage: 1994, 1998a.

\vspace{6.5 cm}

{\bf 45. Pulsar\, Radio Emission}\, reaches\, brightness\, temperatures\, of 
order $10^{30}K$, i.e. is brighter by more than 10 orders of magnitude than any 
other non-manmade radiation in the Universe.
Various attempts by various theorists to explain its (coherent) mode of generation have been proven false by Don Melrose 
(1991). My own first attempt, in 1993, in collaboration with Hsiang-Kuang 
Chang, was based on the model of Reinhold Schaaf \& myself, and on the 
Lorentz-Dirac equation of motion; but we falsely dismissed gyrations -- as one 
of the two linearly independent solutions of the homogeneous equation -- and 
obtained the right power in the wrong frequency regime. Hsiang-Kuang rejected 
our (MAIDER) explanation in his 1994 ph.-d. thesis.
In August 1995 I resumed above approach, and convinced myself that a `MAFER' should do: a microwave amplifier by forced emission of radiation in which the relativistic electrons and positrons, escaping in bunches from a pulsar's polar caps 
along the `open' magnetic field lines, are excited to gyrate coherently around the lines once the field strength has fallen to $\ga 10^7G$. The gyrations are damped on their oscillation time scale, i.e. are extremely short-lived; they constitute a resonance between gyration and small-pitch-angle synchro-curvature radiation, in which the (strongly coherent) radiation term of the Lorentz-Dirac 
equation is (not tiny but) comparable to the gyration term: 1998a. This is the only case I know where Dirac's (third-order, non-linear) radiation recoil term influences the electrons' motion in a significant way. As an aside, Chang \& I (1993) could show that a uniformly accelerated charge does not radiate.

\medskip

{\bf 46. Pulsar Proper Motions} have been thought to be of order $\la 10^{2.5}$
$ Km/s$, for over ten years, reaching some 500 Km/s in exceptional cases only. Such large peculiar velocities (compared with those of their progenitor stars) could have been acquired 
during their formation in a supernova event, both by a liberation of binary kinetic energy (during the sudden mass loss from the system) and by the recoil of an asymmetric explosion, most likely not due to neutrinos (alternative 52) but to different 
magnetic fields in two opposing hemispheres. (The often-quoted rocket effect of an asymmetric radiator, after Harrison \& Tademaru, is a quantitative flaw: radiation has insignificant inertia). Such velocities would still keep the pulsar population bound 
to the Galactic disk, pulsars being alive only during their first runaway, and would not completely erase their birth sites inside the dense-cloud layer of the Galactic spiral arms. This wisdom has been recently challenged -- launched via a few proposed 
pulsar-supernovaremnant associations and corroborated via a rescaling of dispersion-measure based distances (by typically a factor of 2 either way) -- in replacing the reported velocities by much larger ones (factors of 2 to 4):  Lyne \& Lorimer (1994); 
which would partially unbind the pulsars from the Galaxy. But above-mentioned associations tend to confuse supernova shells with pulsar nebulae, and can be re-explained with quite moderate peculiar velocities: 1992, in collaboration with Hsiang-Kuang Chang.
The evidence thus shrinks to quite few `fast stragglers' whose distances may have been overestimated, like the guitar-nebula pulsar: Cordes et al, 1993. More reliable distance estimates are wanted.

\newpage

{\bf 47. Atmospheric Superrotation}, in particular near the equator -- at significant fractions of the speed of sound -- is a phenomenon encountered almost ubiquitously in the solar system: on the Sun, rigidly on Venus, mildly on Earth, most pronouncedly 
on Jupiter and Saturn, and with the (seemingly) `wrong' sign also on Uranus and Neptune. To be more specific: the atmospheres of the outer planets (at least) show both super- and sub-rotation, alternating in latitude belts, with great coloured whirling 
spots at latitudes of maximal shear flow. The outer planets have many moons, moonlets, and dust rings. Are all these phenomena related? And: how in particular can equatorial regions acquire superrotation, opposite to what would result from any type of 
redistribution, ordered or turbulent? Whereas standard `explanations' invoking internal mechanisms (such as convective cooling) face problems: 1983, my own preference is for external torques, either thermally induced tidal (for Venus and Earth: 1977a), or magnetic (for all the others: jointly with Gunnar L\"uttgens, 1998). Magnetic torques are thought to couple the 
conductive interiors of the planets to the solar wind, and are modulated by stick-slip interactions with both their ring systems, moons, and ionospheres. (This coupling has opposite signs for interactions with plasma orbiting inside or outside the corotation distance). The Sun with its multiple systems of torsional oscillations will be discussed in the next alternative. As an aside: differential rotation may well be the feeding mode of most magnetic dynamos, with the possible exception of Ganymede.

\medskip

{\bf 48. The Solar Magnetic Flux} oscillates with a long-term average (Hale) period of 22.2 years, with distinct, non-sinusoidal signatures of its various magnetic multipole moments: 1993b. These oscillations show up not only in daily magnetograms but also 
in (i) sunspot cycles, (ii) aurora cycles, (iii) tree-ring deuterium cycles, (iv) line and continuum radiation (10.7 cm, neutral iron, X-rays), (v) vibrational p-modes, (vi) magnetic multipoles, and (vii) wind strength. They also correlate with torsional (solar surface) oscillations, both of even and odd parity, whereby different tracers (at different heights, or different couplings to the 
magnetic field) rotate differentially: 1992a. Even the same (radio) tracers have different short-term ($\la$ 4d) and long-term (P$_{spin}$) angular velocities. This complex system of motions and the fact that the magnetic field is concentrated into thin, isolated flux tubes cannot easily be described by the linearized `dynamo equation' which assumes that the Sun's 
convection zone regenerates its flux once every eleven years, even though it has had a great deal of success, in particular in describing the `butterfly diagram' of sunspots: Krause \& R\"adler (1980). As stressed by Ron Bracewell and  Robert Dicke, the 
long-term stability of the Hale period points at a `flywheel', or `chronometer' deep inside the Sun, most likely a permanent solar flux frozen into its core which is periodically `modulated', not `generated' by its convection zone. 
Quite generally, a build-up of large-scale (magnetic) structure from
small-scale structure runs counter the Second Law.

\vspace{5 cm}

{\bf 49. The High-Velocity Clouds (HVCs)} in the Galactic halo are discrete neutral-hydrogen clouds `raining' down into the Galactic disk at typical speeds of $10^{2.1 \pm 0.3}Km/s$, intermediate between galactic turbulence and free-fall from infinity, of 
 integrated mass rate $\ga M_{\odot}/yr$. Many of them are associated with (small) molecular `intermediate-velocity' clouds (IVCs) whose frequency of occurrence suggests a collisional origin of the HVCs with some braking medium. What is their origin? In 
1987a, I suggested a similar origin to bubble tracks in a Wilson chamber, or to water droplets condensing on grass stalks over night: the HVCs should have formed from `evaporated' galactic gas, condensing on the (dense, cool) channel walls of the Galactic 
twin jet (from a past Seyfert stage of our Galaxy); see also: 1990c, 1992c, 1996e. During their freefall back into the disk, they are transiently braked by the (presently) feeble, relativistic jet, whence the IVCs. The `evaporation' of the required halo 
gas may happen steadily via cosmic-ray driven `chimneys' above large H II-regions, the cosmic rays dragging along hot interstellar medium (like planetary storms drag along sand): alternative 65. To my great disappointment, this simple explanation has not 
been (shown to be wrong, or) ardently supported by my colleagues at Bonn.

\medskip

{\bf 50. Sgr A West}, the triskelian-shaped thermal radio source which surrounds Sgr A* -- the solar-system sized, non-thermal radio source at the rotation center of the Galactic disk -- projects onto Sgr A East, an even larger, non-thermal radio source 
vaguely resembling a young supernova shell (but smaller and more energetic). All three radio sources are essentially unique in our Galaxy, in strength, morphology, and spectrum: 1990c, 1996b. Are they dynamically related, perhaps like similar sources at 
the centers of `active' galaxies? If so, they should be contained inside of each other, being powered by an unidentified, pointlike `monster' at the center: the burning, rapidly rotating central Galactic disk, which fills them with pair plasma, alternative 
15. Anantharamaiah et al (1991) have concluded against, in showing that Sgr A East is seen in absorption against Sgr A West, i.e. that the latter (and with it Sgr A*) lie in front, not inside of it. But as argued in 1990c, `absorption' can be mimicked by 
lack of emission, whereby the implied low-frequency (Razin) spectral cutoff decreases monotonically in the three sources, from the center outward. The working mode of the Galactic Center is not easy to unravel.

\vspace{6 cm}

{\bf 51. The Mass of Sgr A*} -- the almost unresolved, non-thermal radio source at our Galactic center which has been introduced in the preceding alternative -- has been controversial for many years, between several $10^6$ and several $10^2 M_{\odot}$. The 
large alternative (of $10^{6.4} M_{\odot}$) was suggested by the observed rapid gas motions at separations of $\la 1 pc$ which showed, however, a non-keplerian dependence on separation: 1990c. The small alternative, first proposed by David Allen \& Bob 
Sanders in 1986 and independently by Leonid Ozernoy, was consistent with the center's comparatively low luminosity and with all the other (relativistic wind) phenomena seen in its surroundings. But Eckart \& Genzel (1996) have proved it wrong, by monitoring 
stellar motions down to separations of $\ga 0.01 pc$ and finding a keplerian increase towards Sgr A* consistent with the large alternative; which may therefore be typical for galactic centers (the Copernican principle being trusted). Even so, this large 
mass need not be the mass of the radio-point source Sgr A* but rather the mass of its embedding disk, of solar-system size ($\ga 10^{14.5}$cm): 1996a. Galactic nuclei may hide their secrets for yet another couple of decades.

\bigskip

{\bf 52. The Supernova (SN) Piston} is the medium that transfers the liberated energy -- and radial excess momentum -- of the collapsing progenitor star's compact core to its envelope, ejecting it at typical speeds of 3\% of the speed of light, or $10^{8.8
\pm 0.3}cm/s$. Supernova spectra reveal various different chemical compositions of the ejected shells, classified as SNe of type I and II, with subtypes  Ia,b,c, II-L,-P,b  which tend to be explained as due to various explosion scenarios: Woosley 
\& Weaver (1986). Instead, the different chemistries and powers may be largely due to the different chemistry and size of the progenitor's envelope, type I coming from helium stars, and blue supergiants yielding less powerful lightcurves than red ones (because 
of a higher initial gravitational binding energy): 1988c, 1990a, 1996c, 1998a. Numerical simulations have so far been unsuccessful in getting a shell ejected because they are (necessarily) oversimplified: they lack the magnetic fields (which transfer a large 
fraction of the collapsing core's angular momentum to the ejected shell: 1976a) and they lack their decay product, a relativistic cavity or magnetized pair-plasma, which has to serve as the (relativistic, inmiscible) piston. Without it, a small fraction 
of the ejected mass would have to transiently store the huge radial ejection momentum, would thereby overheat, cool via neutrino losses, and recollapse to form a black hole. Only a `soft' momentum transfer, with transient velocities not much larger than 
the final ones, can lead to an explosion. This quasi-analytical result may be compared with a collision problem of balls whose outcome can be predicted, independent of details, by using the conservation not only of energy but also of momentum. (Several 
of my expert friends have tried to convince me that momentum had no place in explosion physics). More details of SN explosions will be discussed in the following six alternatives.

\vspace{5 cm}

{\bf 53. Supernova Explosions} can be the birth events of neutron stars, as is evidenced by the Crab, Vela, SS 433 (inside W50), and by more than a dozen further `associations': 1998a. The birth of a neutron star should liberate its (huge) gravitational binding 
energy, some $10^{53}erg$, hence should not go unnoticed even if most of the energy escapes in the form of neutrinos: 1985c. We thus expect as many supernovae as birth events of neutron stars -- both pulsars and (binary) non-pulsars -- plus black holes. In 
our Galaxy, birthrate estimates yield one neutron star on average every $\ga 10 yr$: 1988c, whereby there is no hint at missing neutron stars, i.e. at black-hole formations. The SN piston seems to have a high efficiency. Of course, such considerations 
cannot exclude the formation of a black hole in rare (very massive?) cases.

\medskip

{\bf 54. Supernovae (SNe)} tend to be described by `strong' (Sedov-Taylor, shock) waves -- like nuclear bombs in the atmosphere -- after Shklovskii (1962). Strong waves are good approximations for thin-walled, or pressure bombs, but fail as approximations 
for thick-walled, or splinter (shrapnel) bombs: the latter don't sweep, hence have a much larger range. In a SN, the energy is liberated in a volume not much larger than that of a neutron star, of radius several $10^6 cm$, and transferred to the progenitor 
star's envelope, of radius $10^{13 \pm 0.5}cm$, i.e. transferred through some seven decades in radial distance. It should therefore not take by surprise that SNe behave like splinter bombs: 1988c, 1990a, 1996c. A more exact reason is Rayleigh-Taylor 
instabilities of the pair-plasma piston during the ejection: in pressure balance with the thermal matter, the latter can only fill a negligible fraction of the volume. Independently, a splinter morphology is recognisable in (the velocity-spread of) SN spectra, 
and again in most SN shells: ringlike-looking shells in particular are the illuminated outer edges of former windzones, i.e. of material swept up by the late progenitor's wind. SNe act 
similarly to flashlights.

\vspace{5.5 cm}

{\bf 55. Supernova Shells} are strong radio emitters. Their non-thermal spectra allow them to be distinguished from `H II-regions', the ionized environs of bright stars. (In many cases, H II-regions and supernova shells occur jointly because massive stars 
are born in clusters, and are short-lived). 
Where do the relativistic electrons (and positrons?) in SN shells come from? An almost `generally accepted' wisdom tells that these relativistic leptons are accelerated `in situ', by a large number of bounces off fast-moving shocks, defying the Second Law 
(like the distribution of richness among people); cf. Reynolds \& Chevalley (1984), but also alternative 11. An extreme case was SN 1987A which flared at radio frequencies within days after its appearance, much too fast for in-situ acceleration of relativistic 
electrons. As already argued in alternative 52, it is my understanding that relativistic leptons are formed in the core of every SN explosion, via magnetic reconnections. The frequent objection that such a relativistic bubble would lose its energy 
by cooling under adiabatic expansion, through $\ga 12$ orders of magnitude in radius, can be invalidated by reminding that (i) the (former) windzone around an exploding star is of such low density that for some time, the pair-plasma explosion cloud can 
expand quasi freely, like in vacuum, and (ii) as demonstrated by the ubiquitous jet sources, when the quasi-weightless relativistic plasma meets with resistance it prefers to ram vacuum channels through which succeeding generations of charges can propagate 
free of expansion losses: 1988c, 1990a.

\medskip

{\bf 56. Supernova Shells} tend to be understood as luminous (strong) shock waves sweeping the circumstellar medium (CSM), cf. Falle (1981), rather than as the structured CSM flaring when traversed by the shell-shaped cloud of ($\ga 10^4$) filamentary SN 
ejecta; (alternative 52). In a number of shells -- like Cas A -- luminous `knots' and `flocculi' can be distinguished differing in (i) speed (by an order of magnitude), (ii) mass (inversely to speed), and (iii) chemistry, to be understood as the ejecta 
and their targets: 1985c, 1995, 1996c. The various sizes, morphologies, and measured  SN shell velocities -- despite a fair uniformity of ejection velocities inferred from SN spectra -- have led to a lot of confusion in the extended literature. Their slowdown is described in my 1984 paper with Gopal-Krishna.

\vspace{7.0 cm}

{\bf 57. Supernova Shells store} a mechanical energy of $10^{51 \pm 0.5} erg$ in radial expansion, rather independently of the ejected mass, but have an integrated light output of only $10^{49.5 \pm 0.5} erg$, some 3\%. Where does the excess energy go? 
Looking at the spectrum of our Galaxy, see figure above 69., I can see the cumulative SN energy escape in the infrared window, emitted by forming (atomic and molecular) clouds. Apparently, SNe heat their environs and make the Galactic disk expand locally 
whilst it recontracts in cloud-forming regions, the clouds emitting (part of) their contraction energy: 1988c.

\bigskip

{\bf 58. Supernova Lightcurves} can in principle be powered by various sources: by (i) the explosion energy itself, tapped both from the heat of the boosted matter, and from crashes of overtaking filaments ejected at different speeds, (ii) radioactive decay 
of the ejected nuclear ashes, (iii) magnetobremsstrahlung of the SN piston (alternative 52), and (iv) cooling radiation of the stellar remnant at the center, most likely a hot neutron star. Sources (i) and (iii) would only involve a small fraction of 
the available energy, some 3\%, whose precise value is difficult to predict; source (ii) would be ignorable unless the SN had produced a huge amount of radioactive $^{56}Ni$ (which decays via $^{56}Co$ to $^{56}Fe$), and unless escape losses of its emitted 
decay $\gamma$-rays were modest; and source (iv) would easily power the light curve if it could escape as electromagnetic radiation (rather than neutrinos), e.g. as cooling radiation from a handful of volcanoes. The literature has largely concentrated on 
the radioactive source (ii), partly because the radioactive-decay $e^{-1}$-folding times (of 9d and 111d, respectively) are comparable to observed $e^{-1}$-folding times of the lightcurves (which vary, however, by factors of $\la2$ from SN to SN), and because simple ejection 
models seem to require an additional energy source: McCray (1993). But radiation transfer through a filamentary shell yields likewise exponentially decreasing intensities, as well as the remarkable plateau in the brightness temperature (at B - V $\approx 0$, 
corresponding to $10^{3.8 \pm 0.1}K$) for about two years after the first fortnight: 1988c, 1990a, 1998a, and an unpublished Bodrum lecture in 1993. Supernovae may behave like the (photon) bags which were missing to the Schildb\"urger.

\vspace{7 cm}

{\bf 59. The `Exotic' Supernova Remnants (SNRs)}, whose morphologies resemble birds, a rabbit, tornado, mouse, puff, or sickle, cannot easily be understood as luminous explosion shells even though their spectra and (radio) energetics resemble the latter.
Quite often they contain a compact core in offcenter position whose presence has seemed to ask for an independent cataclysmic event. But as first explained in 1992b and further elaborated jointly with Hsiang-Kuang Chang: 1992, the exotic SNRs may all be 
powered by young pulsars (PSRs) rather than by the (pair plasma from the) explosion that gave rise to the PSR; whereby the PSR surrounds itself with a luminous, fast-expanding bowshock of much smaller extent (R = $10^{17.3 \pm 1}cm$) than the complete SNR, 
or rather `Pulsar Nebula'.

\bigskip

{\bf 60. The Fireworks in Orion} has entered conservative text books like Uns\"old \& Baschek (1988) as a bipolar, protostellar wind emerging from the infrared Becklin-Neugebauer or Kleinmann-Low object, an interpretation which has found seeming support 
by a number of numerical simulations in recent literature. In collaboration with Aylin Yar, I have convinced myself by 1995 that we deal with an unrecorded supernova, some $10^{2.2 \pm 0.2}yr$ ago, of which we only see the tail of the cloud of ejecta; cf. 
1995, and Kundt \& Yar (1997). The SN interpretation is indicated by a perfect Hubble-flow: $\vec v \sim \vec r$, on scales between $10^{-2}$ and 0.6 pc, as well as by the detailed shrapnel morphology revealed by high-resolution IR maps, cf. alternative 52.

\vspace{7 cm}

{\bf 61. The `Fossil' Fuels} natural gas, oil, and coal are commonly understood as due to buried former swamps, or peat fields whose fluid and gaseous reaction products accumulate in large underground basins capped by impervious rock, and whose older and 
solid remains form the various coal layers, or `beds' -- lignite, coal, and anthracite -- increasingly enriched in carbon with age. Evidence of their `biogenic' origin are traces of bacteria which once produced the specific organic compounds, variable from place to place. This conservative explanation offers no simple answer to why there are beds of (almost) pure anthracite as thick as 300 m (!), why gas, oil, and/or coal tend to be found in alternating layers essentially everywhere on Earth, in various 
successions one above the other, and why all the hydrocarbons hide from sight which were once contained in the carbonaceous chondrites, at the 5\% level, which formed a significant fraction of the meteoritic building blocks of planet Earth. Instead, I have 
learned from Thomas Gold in 1982, (cf. Gold, 1987), that a much more convincing explanation of the origin of the natural fuels is an essentially abiogenic one: organic material -- buried e.g. by volcanic ashes -- may serve as a porous layer which is vented, 
throughout millions of years, by methane and similar abiogenic gases from deep below, on their way to the surface. Underground bacteria, enjoying somewhat higher temperatures and pressures than prevail near the surface, digest the methane and convert it to 
either oil or coal depending on the ambient conditions, in particular on the abundance of water 
(Frederickson \& Onstott, 1996). Coal is mostly amorphous, but can contain perfectly structured organic inclusions. In particular, former tree trunks can be found strongly enriched in carbon; 
but also -- under different conditions, in the petrified woods -- strongly enriched in silica; in neither case do they form the main substratum, only the nucleation site of the deposit. This alternative (of Gold's) has stimulated my work with Axel Jessner 
(on volcanism): 1986. It has been the bitter, life-long controversy between
William Plotts (1940) and his generation.-  The escape of natural gas from deep below can perhaps be as violent as has been the Tunguska catastrophe on 30 June 1908 which, for several decades, was (erroneously?) believed to be of (stony) asteroidal origin.
\\ .

\vspace{0.5 cm}

{\bf 62. Plate Tectonics on Earth} have been detected by Alfred Wegener, by seeing the eastern coastline of South America match the western one of South Africa (as well as their rock compositions, and lifeforms), but has been denied by Harold Jeffreys because 
of the huge frictional resistance experienced by a plate whose diameter measures thousands of Km, and thickness over 70 Km. In a sense, both of them were right. Wegener has been shown to be right by measurements of the plate-drift velocities, both indirectly, 
by the varying magnetizations of ejected and solidified magma, and directly, by modern VLBI: the plates move at average speeds of several cm per year. But Jeffreys was likewise right in pointing out that the known (thermal) crustal forces fell short of pushing the 
plates around, by much more than an order of magnitude. Thermal convection rolls, as are usually considered, cannot do the job. In my 1986 communication with Axel Jessner, we argue that deep-rooted chains of volcanoes in statu nascendi, so-called `volcanic fences', can exert high enough pressures to make the plates move, like wooden wedges which can split a rock. More in detail, we argue that volcanic pipes form via a natural instability of a solid on top of its melt, by overhead stoping, and can exert enormous pressures near their (covered) tops when the (gas-enriched) melt is lighter than the ambient rock, because of equal pressures at the (fluid) bottom. Linear arrays of such overpressure pipes will form `rift systems', or `ridges' (like the mid-Atlantic one) which define 
plate boundaries, and can be secularly stable once a few pipes have forced their way up, and started to drive the adjacent plates apart. Such driving apparently happens stripewise in $\la m$-sized steps, once every few decades, the mapped `transform faults' being the edges of such stripes. At the time of our writing, volcanism was considered `shallow'-rooted, $\la 700 Km$. When I saw Strobach (1991) draw the volcanic pipes feeding the ocean-island volcanoes (over `hot spots', like Hawaii) all the way down 
to the molten core, and looked at the results of chemical analyses of their ejecta, I jumped to the conclusion that plate tectonics is likewise rooted as deep as the Earth's molten core, some $10^{3.5}Km$: 1991b.

\vspace{6.7 cm}

{\bf 63. The Earth's Magnetic Field} is anchored in its fluid core, as has been strongly suggested by satellite measurements of its first 29 multipole moments (by MAGSAT): The magnetic moments decrease exponentially with multipole number, corresponding to 
roughly equal energy densities at a depth of $(3050 \pm 50) Km$, some 150 Km below the surface of the molten metallic core. On top of this simple behaviour, the moments show secular equipartition (white-noise) variability both inside the core, and near 
the surface, the latter evidenced by the multipole moments above the $14^{th}$. Independently of these MAGSAT results, the Earth's magnetic-field anomalies are known to drift westward, at rates of $\la 0.3^o/yr$, implying that the core of Earth spins more 
slowly than its mantle -- contrary to standard wisdom that the mantle of Earth is braked by both lunar and solar tidal torques, most strongly at the shallow edges of the ocean basins. In 1989, Hans Volland and I resolved this seeming paradox by pointing 
out that magnetic friction of the Earth's magnetosphere on the solar wind can decelerate its anchoring core more strongly than tidal forces decelerate the mantle. The required (strong) magnetic torque is supported by observed cometary-tail accelerations, 
considered earlier in collaboration with Vinod Krishan: 1988. This re-interpretation of the Earth's spin motion implied that the typical electric conductivity $\sigma$ of its mantle had to be much lower than assumed earlier, $\sigma \la 0.1 S/m = 10^9 s^{-1}
$, and that a proposed explanation of the decadic fluctuations of the LOD by core-mantle coupling was untenable (next alternative). We noticed only afterwards that Lay (1989) had just measured the conductivity of perovskite under mid-mantle conditions, 
and found $\sigma = 0.01 S/m$.

\medskip

{\bf 64. The Length-Of-the-Day (LOD)}, or spin period of the Earth's mantle, can be monitored with an accuracy of at least 0.02 msec. On top of a secular increase by (1.7 $\pm$ 0.1) msec/century , and of periodic annual and seasonal oscillations, it shows 
 long-term fluctuations of amplitude $\la$ 2 msec, the latter known as the `decadic' fluctuations. Whilst the former can all be understood as angular-momentum exchanges between the mantle and the atmosphere ($\la 0.4$ msec) plus ocean currents ($\la$ 0.05 
msec), the latter have been attributed to angular-momentum exchanges with the fluid core. In my 1989 work with Hans Volland discussed in the preceding alternative, we have ruled out such an explanation. How constant is the mantle's moment of inertia, (depending on glaciers, vegetation and ground-water distribution, and acted upon by solar and lunar tides)?

\vspace{6.5 cm}

{\bf 65. The Galactic Disk} is commonly thought to be filled with `warm' ($10^4K$) interstellar matter (ISM), mainly hydrogen and helium, but Ron Reynolds does not deduce more than 0.2 of the required column density from his H$\alpha$ measurements. What 
fills the Galactic disk? For comparable pressure and lower (column) density, a volume-filling component should be hotter (than warm) in order to fill it. UV- and X-ray maps show strong inhomogeneity (unexpected for a volume-filling component), and again 
insufficient column density. For these reasons, I have considered (relativistic electron-positron) pair plasma as a candidate to fill the Galactic disk: 1992c. As argued above (32.,49.,51.,55.), expected sources are (i) all (sufficiently magnetized and 
spinning) compact stars, in particular neutron stars, (ii) all supernova explosions, and (iii) (the central engine) Sgr A*. For a mean pair-escape time of $10^{7 \pm 0.5}$yr from the disk, like that of the cosmic rays -- or rather for a mean annihilation time 
of $10^{5.5}$yr, corresponding to the observed $10^{43}$ annihilations per sec -- these sources should be able to steadily replenish the losses. The (magnetized) pair plasma would hardly mix with the ISM, so that bremsstrahlung $\gamma$-ray emissions 
constitute only a boundary-layer phenomenon, not a volume phenomenon (like inverse-Compton radiation). Their energy density would be more-or-less comparable to that of the ionic cosmic rays. The escape (of some non-annihilated 3\% of them) from the disk would 
happen like that of the ionic cosmic rays: through several hundred Galactic chimneys, issuing from high-pressure H II-regions, of which `Stockert's Chimney' is the (only well-known) prototype, as argued in 1987, in collaboration with Peter M\"uller. The 
often-stated impression that the solar system finds itself inside a `local hot bubble' is then understandable in analogy to one's being inside a forest where the near environment is transparent whereas tree trunks bar the view in all directions beyond some 
`covering' distance: just replace trunks by (striated) H I-clouds. At the same time, we understand why the Galactic disk is a strong emitter ($10^{37}erg/s$) of the 511 KeV pair-annihilation $\gamma$-ray line.

\vspace{0.5 cm}

{\bf 66. Accretion Disks} occur abundantly in astrophysics: (i) the Milky Way (as the prototype of galactic disks, which accrete onto their central AGN engine), (ii) accretion disks in X-ray sources, which feed white dwarfs, neutron stars, or possibly black 
holes, and (iii) accretion disks in star-forming regions, the progenitors of multiple-star and planetary systems. A common property of all of them is their non-stationarity: Differential rotation plus viscosity give rise to angular-momentum transport 
(radially) outward, equivalent to their mass spiralling inward, in direct proportion to friction. What is the dominant cause of a disk's viscosity? A reliable answer should model their complete dynamics, including their warping, caused, among others, by a tilted central magnetic dipole (as treated in 1980 with Marko Robnik, and in 1989 by Susanne Horn and myself). But even in the absence of external forces, turbulence is a ubiquitous cause of viscosity, orders of magnitude larger than molecular viscosity, 
and is generally held to be the dominant reason for a disk's viscosity. Turbulence may or may not be helpful in magnetizing a disk, cf. 1993b. In any case, there is a strict upper limit to what turbulence can achieve -- often expressed by the so-called viscosity 
parameter $\alpha$ being less than unity -- which can be lower than what is required to drive a disk's spiral-in motion. In such cases, magnetic tensions (of a predominantly toroidal field, exerted over a significant height) can be found to yield the torque necessary for the estimated 
mass accretion rates: 1990b. Disks can evolve through magnetic tensions.

\newpage

{\bf 67. $\gamma$-Ray Bursts} reach the Earth at a rate of three per day on average (at present-day sensitivities), with photon energies ranging from X-rays up to almost GeV energies -- in extreme cases even up to tens of GeV -- with spectral peaks logarithmically
equi-distributed between less than $10^2$ KeV and more than 1 MeV, during time intervals of $10^{0.5 \pm 2}$sec but with occasional tails lasting for as long as $\la 10^2$min, and with temporal fine structure down to $\ga$ msec and shorter. What are 
their sources? For over a decade, the bursts had been thought to be emitted by old Galactic neutron stars, accreting in transit through Galactic clouds, and radiating briefly at almost their Eddington rates. This interpretation was strengthened by occasional 
reports of periodic modulations, of X-ray line features looking like cyclotron, and of $\gamma$-ray line features looking like redshifted pair annihilation. Such reports have not been confirmed recently; instead, better event statistics found isotropic arrival directions from a region of limited range, which has influenced model builders to explore source locations at either cosmic, or at least distant-halo distances. Instead, in my 1993 work with Hsiang-Kuang Chang we defend the more 
conservative scenario of nearby ($\la$ 0.5Kpc) Galactic neutron stars accreting spasmodically and emitting the bursts in a highly non-isotropic manner, as transrelativistic sparks from slightly above their surfaces; whereby an increasing source number with distance is first-order compensated by a decreasing probability of being in the beam. (See also alternative 79 for terrestrial $\gamma$-ray bursts).

\medskip

{\bf 68. The Soft $\gamma$-Ray Repeaters (SGRs)}, so far three in number, are $\gamma$-ray burst sources of the kind just described but from whose direction bursts have been recorded repeatedly, many times per year, with somewhat softer spectra (except for one very 
strong burst), and with occasional  periodicities. They have been identified with neutron stars inside of supernova remnants, with a companion of an ultraluminous star, and/or with a radio-jet generator. None of their distances are reliable. Hsiang-Kuang 
Chang \& I (1994) have interpreted them as nearby neutron stars blowing pulsar nebulae 
(alternative 59) -- the nearest among all $\gamma$-ray bursters -- at distances of $\la$ 50 pc.

\vspace{4.5 cm}

{\bf 69. The 2.73 K Background Radiation} has yielded the Nobel prize for its discoverers, and has been interpreted by most as a relict from a hot, early stage of the expanding Universe which decoupled some $10^{10}$ years ago -- with only a few sceptics, 
among them Fred Hoyle (1975) and David Layzer (1990). A worry for the hot-big-bang interpretation is its remarkable blackness: T = 2.728 K, despite the fact that (re-) combination and decoupling of the cosmic plasma should have been terminated by a large 
number of Ly-edge re-absorptions from the Wien branch followed by fluorescent re-emissions which overpopulate the Rayleigh-Jeans branch, thereby destroying blackness. Instead, Layzer prefers (the logically simpler case of) a cold bang which implies a different 
predicted rate of chemical-element (He) production during the first three minutes, depending on the lepton-to-baryon ratio, and avoids the problems of the hot bang in explaining the formation of cosmic structure (from the bottom up), but has not yet 
presented a convincing explanation of how a postulated, fast-burning first star generation emits the going-to-be cosmic background radiation and finds enough scatterers to guarantee its universal blackness. At this point, I wonder whether the evaporating 
low-mass branch of the (disrupted, solid) primordial hydrogen-mass spectrum can serve as the scatterer -- during its final evaporation stage -- in the form of hydrogen snow. The high-mass branch would have served as the first-star generation, and the intermediate-mass branch as the (`missing', baryonic) cold dark matter in the Universe.

\medskip

{\bf 70. The oldest (Galactic) Stars} known, often used to bound the observed age of the Universe from below, are those in globular clusters leaving the main sequence. Their age is estimated by various experts on stellar evolution at  t $\ge$ (17 $\pm$ 3) 
Gyr, cf. Chaboyer et al (1996), in marginal conflict with the present best estimate of the world's expansion (or Hubble) age (for a vanishing cosmological constant), t $\la$ 2R/3$\dot R \la$ 10 Gyr. Is the conflict real? No conservation law can be invoked for the inequality. At this point I wonder how reliably 
one can estimate the burning time of a $\ga$ solar-mass star to the (ill-defined) point where it leaves the main sequence, in particular given the unsolved problem with the solar neutrinos; cannot (17 $\pm$ 3) be as believable as (14 $\pm$ 4) ?

\newpage

{\bf 71. Superhumps} in the lightcurves of cataclysmic variables, best known in dwarf-nova `super-outbursts' during the SU Uma stage (of short-period X-ray binaries containing a white dwarf), are periodic modulations of the large ($10^{2.2 \pm 0.6}$-fold) transient rises of the optical output. During the plateau-shaped super-outbursts, the intensity is weakly and 
periodically modulated in a spiky, i.e. non-sinusoidal way, with a period that is longer than the orbital period by (4$\pm$4)\% for short-period binaries [$P_{orb} \la 3(M/0.5 M_{\odot})$h for total mass M] -- and shorter for long-period binaries -- and that can still be retrieved shortly after the outburst, but with a phaseshift of half a cycle. The outbursts are interpreted as caused by sudden increased mass ejections by the donor star, and their modulations as somehow due to the flaring accretion disk; but the latter interpretation is not convincing. A similar phenomenon has been encountered 
recently in the class of transient bright low-mass X-ray binaries -- mostly interpreted as black holes, cf. alternative 27 -- whose optical lightcurves have shown superhump-like 
periodic modulations during outburst (in the case of Nova Muscae: Remillard et al, 1992). In both cases, my preferred interpretation of the lengthened (shortened) period is a modulated illumination of a string of `blobs' of ejected matter orbiting the binary system slightly 
outside (inside) the Roche lobe: 1996d, 1998a.

\bigskip

{\bf 72. The Supersoft (X-ray) Sources (SSSs)} are a class of some $\ge$34 bright, very soft X-ray sources in the LMC, SMC, Andromeda galaxy, and our own discovered by ROSAT, radiating near the Eddington luminosity of a solar mass, some $10^{38}$ erg/s, with a spectrum 
peaking between 20 and 60 eV. The softness of their spectra suggests that white dwarfs could be involved, but even white dwarfs tend to radiate at harder photon energies; and occasional anticorrelations between X-ray outbursts and the optical light 
curve indicate that the optical emission may come predominantly from an illuminated windzone, inconsistent with a white-dwarf interpretation already because of its emission size. The class also contains neutron-star binaries like SMC X-1. For these and other 
reasons, I have convinced myself that the known multiple lightcurves ask for neutron-star binaries with massive accretion disks, with the supersoft X-rays coming from the disk: 1996d. More in detail, the low Eddington rate of $\la 10^{-8}M_{\odot}/yr$ 
for a neutron star can lead to epochs during which its donor star transfers supercritically, and fills up the disk beyond the test-mass regime, to a mass of one or several $M_{\odot}$. (For white dwarfs this situation hardly occurs, because of a $10^3$-times 
larger critical accretion rate). During its growth, the disk should be a bright, supersoft X-ray source. A heavy disk may also transiently give rise to a super-Eddington (X-ray) source, i.e. to a source like SMC X-1, LMC X-3, LMC X-4 and $\ge$ three other neutron-star binaries whose massive feeding can apparently overcome the repulsive radiation pressure known to prevent super-Eddington accretion in near-spherical geometries. Last but not least, neutron-star binaries with self-gravitating accretion disks may well explain all the properties of the (over 15) `black-hole candidates', alternative 27, in particular (in the low-mass case) their occasional decade-long quiescent intervals, and nova-like outbursts at $\la 10^{38.5}erg/s$ : 1996c, 1998a.

\vspace{6.5 cm}

{\bf 73. Spermwhales} feed at the bottom of the ocean, on squids, at depths that can exceed three Km, but they have to get back to the surface for breathing. Their deep diving is typically done in 1.5 hours, at speeds outrunning submarines. Do they have 
to perform work -- like almost all other animals (excepting a few further sea mammals, like bottlenose, sea elephant, and walrus) -- to move up and down between different gravity levels? When I first expressed my belief that they would get all their shuttling 
essentially for free, in 1992, led by the conviction that this unusual routine would otherwise not have been compatible with more sedate habits, I met with scepticism by both friends and professionals. But then I found out the many devices in which 
spermwhales differ from other animals, cf. Denny (1993): (i) They are well insulated, like other animals living in cold climates, by fat under their skin; but they can use their fins for cooling, mediated by blood circulation through them, arteries and veins 
being in counter-current array near the skin. During deep diving, blood circulation is interrupted, via valves, the blood being stored in the `Wundernetze'; only heart and brain are supplied by a reduced circulation, so that heat losses are minimal. (ii) 
Oxygen is largely ($\la$ 50\%) stored in muscles and tissues, bound to myoglobin; spermwhales therefore exhale at the beginning of diving, to get rid of the nitrogen; the lungs thereby collapse. (iii) The huge volume of wax in its head that has given 
the spermwhale its name, and the oil along its backbone, are almost incompressible but have a $\ga$ 10 times higher thermal expansion coefficient than water; the wax melts, and expands near $36^o$C, the temperature above which diving comes 
to an end. (iv) Before deep diving, the animal inhales, and cools for over 15 min until the wax has frozen, and its head gets heavy. The spermwhale's body temperature rises during diving, in proportion to the burnt amount of oxygen, and with it its buoyancy 
(by at least $10^{-3}$), thus controlling the epoch of `taken breath' reliably like a sand clock.

\medskip

{\bf 74. The Na$^+$-K$^+$ pumps in Cell Membranes} generate a cross\,\-membrane voltage of 0.07 V, powered by ATP; they are the universal electric generators in animals, achieving voltages of up to 0.8 kiloV in the extreme case of an electric eel, by being stacked in 
series: Alberts et al (1989). The pumps allow a cell to feed on mesoscopically large resources (like starch molecules), which are dragged across a sluice (`symport') in the enclosing membrane by being electrically charged with a sodium ion. They work diffusively on the msec timescale, thereby 
setting a lower bound on biological reaction speeds (when charging nerves). How efficient is the action of the pumps? When Marko Robnik and I tried to understand the functioning of the Na$^+$-K$^+$ pumps, stimulated by David Layzer's Cosmogenesis (1990) which left the true mechanism 
unexplained, we noticed that very likely, these ion pumps work like heat pumps: a voltage of 0.07 V can be overcome by the (stochastic) heat motion of a sodium ion at body temperature, hence all that is required is an efficient gating (by the ATPase) 
that prevents back flow, down the potential gradient; the gating must be actively powered, but most of the electric energy can be gained by a (slight) cooling of the environment, as a heat pump, thereby reaching a high thermodynamic efficiency, of 66\%. Our 
1994 communication to `Science' was rejected without an explanation.

\vspace{6.5 cm}

{\bf 75. Navigation via Magnetic Fields} is often thought to be done by migrating birds like doves, because no other mechanism of long-distance orientation is apparent; but the magnetic organ has not been found by 1996. (Short-distance navigation should be based on smell, 
which is more directly suited for survival). Instead, several species of fish are known to orient electrically -- via self-generated pulses -- and others to orient magnetically via the Lorentz force  e $\vec \beta \times \vec B$  whose electric field they can supposedly sense down to a level 
of 5 nV/cm. Once fish can sense $\vec B$ when swimming at an angle to it, birds at their much higher (flying) speeds should find it much easier to sense it via the induced Lorentz field, perhaps by having a thin electric conductor running across their head 
(or beak?), with a sensitive voltmeter placed in the middle. Such sensitive nerves threading magnetite crystals have indeed been detected in the upper beak of homing pigeons in 1997 by Gerta Fleissner and Elke Holtkamp-Roetzler (Frankfurt).

\vspace{0.5 cm}

{\bf 76. Plants show Exudation, and Root Pressure} of up to 6 bar which cannot be explained by capillarity, already due to its sign, and not by osmosis either because a reverse osmosis is involved, across the endodermis cell layer in the outermost root-hair 
zone of (young) root tips where a high osmolarity of the cortex -- needed to absorb the water from the soil -- is reduced to near its ambient value in order that the upper parts of the plant can lift the sap (again) osmotically: e.g. Nultsch (1991). Even though transpiration 
is the motor that propels the necessary circulation of the sap in plants under favourable conditions, exudation cannot be dismissed either: it plays a similar r\^ole to the starter of a car, helping the motor `transpiration' whenever it has been transiently 
turned off, or reduced, by driness of the soil, by darkness, or by high air moisture. How is root pressure generated? The answer at which I have arrived during over four years, starting in 1992 jointly with Marko Robnik, are cell-sized mechanical pumps whose rigid cases are 
the Casparian-girdled endodermis cells, whose 1Hz pistons are the folded outer endoclinic cell walls, and whose valves are the large number ($\ga 10^3$) of plasmodesmata in the many `pits' of that wall which succeed in reducing the osmolarity by their 
mesoscopic narrowness, steered each by an ATP-powered `sphincter' and by the dumb-bell-shaped endoplasmic reticulum (ER). The valves achieve a reduction of the concentration (at nearly unreduced pressure) at the expense of thermal energy (equivalent to $\la$0.2K); i.e. heat pumps are in operation in 
plant roots -- as in the Na$^+$-K$^+$ pumps (alternative 74) -- physically required to do the reverse osmosis of the transiently overconcentrated sap. 
After eight leading international journals have "found themselves unable" to publish our manuscript, Vadim Volkov came at a rescue: 1998; see also 1998b.

\newpage

{\bf 77. Photosynthesis} is the universal mechanism by means of which plants transform solar energy into chemical energy, in the form of ATP (= adenosin-tri-phosphate), of NADPH (a reduced form of NADP = nicotinamid-adenin-di\-nucleo\-tide-phosphate), and subsequently 
of starch: e.g. Nultsch (1991). Despite impressive insights that have been gained into its functioning, a complete understanding of photosynthesis is still lacking. Apparently, electrons inside chlorophyll molecules are excited by the photoelectric 
effect, and cascade down a chain of overlapping bound states to the outer edge of the thylacoid membrane -- steered perhaps by cis-trans isomerisations -- whilst protons on the inner edge of the membrane are set free to move along its surface. In this way, a (large!) membrane voltage is generated by (cascading) 
bound electrons, and discharged via the protons which close the current by falling through membrane channels at whose ends they synthesize ATP. Very likely, this ADP $\rightarrow$ ATP conversion makes use of the fact that at the same voltage, protons 
have a $10^{3.3}$ times larger momentum than electrons.

\bigskip

{\bf 78. Water in the Solar System} is of importance for life and for an exploration of space 
(Dyson, 1985). It is abundant on Earth, on the comets, and probably on most of the solar-system bodies as well, but in what quantities? In 1986, Louis Frank claimed to have detected occasional short-lived ($\approx$ minute) `holes' in UV images of Earth (illuminated by the Sun, the images made by 
spacecraft), of size $\ga$30 Km; see Frank \& Huyghe (1990). The UV holes occurred at a rate of three per minute when extrapolated to the whole Earth, and were supposedly confirmed in 1997, at a much higher significance level. Frank has interpreted them as caused by infalling house-sized objects made of frozen water whose typical mass was dictated by their ability to absorb the 
dayside UV glow throughout the hole. This interpretation aroused criticism because an infall rate of three per minute is characteristic of g-sized objects -- according to the inner-solar-system 
power-law distribution of orbiting masses: M$^2 \dot{N}_M \approx$ 10$^{-20.5}$ g cm${-2}$ s$^{-1}$  for 10$^{-18} \la$ M/g $\la$ 10$^{18}$ , known to hold from impacts on Moon and Earth. I therefore re-interpreted Frank's UV holes in 1987 (unpublished) as caused by `ice cherries' 
which hit the upper ionosphere -- at heights of $\ga$ 0.3 Mm -- in the form of house-sized vapour clouds, and transiently blow a tens-of-Km wide vacuum channel which recollapses on the timescale of a minute. Such a heavenly hail would supply water vapour to the atmosphere 
at a rate of some 10${-9}$ times that of volcanic outgassing -- i.e. would not be responsible for the oceans on Earth -- but would be consistent with recent indications of some 10$^{14 \pm 2}$g of ice on the Moon's polar caps (where they may have accumulated by sublimation; Reichhardt, 1998), 
with the findings of water vapour in the upper atmospheres of the outer planets
-- distinctly more than predicted by the barometric-height formula: Hunten
(1957) -- and with a detected ``icy grain halo'' of comet Hyakutake (Harris et
al, Science, 1. 8. 1997, 676 - 681). 

\newpage

\noindent{Water -- the universal 
transport medium of life -- may pervade the whole solar system.}

\vspace{6 cm}

{\bf 79. Planet Earth is charged} negatively to 0.4 MV w.r.t. its ionosphere, with a daily modulation of $\ga$ 10\%, peaking at 19$^h$GMT, i.e. during the late afternoon of the African and European continents. Differential thunderstorm voltages can be higher -- up to 0.1 GV -- resulting in the buildup of ionized discharge channels: lightning. Who charges whom? Whereas 
 the literature of the second halfth of this century holds the thunderstorms responsible for charging the atmospheric `condenser', Gernot Thuma \& I (1998) have convinced ourselves that negative charging takes place all the time, in the fairweather atmosphere, via a steady drizzle of its 
negatively-charged `heavy' aerosols, of radii between 10$^{-1}$ and 10$^2 \mu$m. Their negative charge is acquired through impacts of electrons, during their transient free epochs after having been kicked loose by (1) radioactive-decay products, (2) cosmic-ray shower particles, or (3) hard solar photons. Most of the compensating positive charge hovers at mid-tropospheric 
altitudes, between 1 and 3 Km. During thunderstorms, winds with vertical components distort and lift the volume-charge layer towards the bottoms of clouds whose subsequent rain transports them back down and to the ground. In this way, and via additional charged rainfalls, both horizontal and vertical charge gradients are locally 
generated, leading to all sorts of lightning. Less frequently, thunderstorm clouds discharge towards the ionosphere, with hot emissions ranging up into the soft $\gamma$-ray range (alternative 67).

\newpage

\section {Non-Alternative Publications}

The preceding seventy-nine alternatives may leave the reader with the impression that my scientific work has been aimed at finding loopholes in the published literature. To be sure, all I have strived for is reaching a thorough understanding of what had 
been achieved. In support of this thesis, I have added seven entries to the list of references which do not qualify as `alternative' explanations but exclusively as attempts at deeper understanding. They are:

(1) The second printed text of my life -- submitted in the spring of 1958 -- complements a theorem proven in the book by Lynn H. Loomis, on closed ideals of certain (commutative) regular, semi-simple Banach algebras. My seminar teacher Ernst Witt had raised the problem, and urged that its 
solution be published. As with a large number of other mathematical problems, Werner Boege had helped me grasp the issue.

(2) My work on exact solutions of the gravitational field equations, reported in 1962 in a joint textbook article with J\"urgen Ehlers, was aimed at exploring the (coordinate-invariant) properties of 4-dim Lorentzian manifolds, used after Albert Einstein 
to describe spacetime geometries. How different are their properties from Newtonian mechanics? Among others, our article contains necessary and sufficient (constructive) conditions for two spacetimes to be isometric -- theorem 2-2.6 -- and finds a surprising 
one-to-one correspondence between a large class of electromagnetic waves and their (non-linear!) gravitational analogues.

(3) In my 1966 habilitation thesis, I made a fierceful attempt at quantizing General Relativity. In comparing the different canonical approaches by Paul Dirac, Peter Bergmann \& Arthur Komar, Richard Arnowitt, Stanley Deser \& Charles Misner, and by Bryce de Witt, I could show their strict equivalence at the classical level but saw no way towards a (unique) quantized theory of gravity. (Embarrassing is the wrong lemma 1 in section 3, in which I confused Lie ideals for single and iterated Lie multiplication). Despite Ashtekar's new variables (1986), I am no longer sure whether fields should be quantized at all: Asim Barut (1988) argues that QED may lead the wrong way, and so does Klaus Hasselmann's metron approach (this volume) of realizing Einstein's dream.

(4) Thanks to Ted Newman, I could spend my honeymoon year in the U.S., and enjoy with him -- among others -- an exploration of the general hyperbolic differential equation in two dimensions. We found large classes of exact solutions not contained in mathematical 
reviews -- though in part known to P.S. Laplace (as it turned out later) -- and criteria of whether or not they develop wave tails. Thanks to F.G. Friedlander, our lengthy analysis was printed in 1968.

(5) I had started my scientific work in the field of General Relativity, because of Pascual Jordan -- "the unsung hero among the creators of quantum mechanics", according to Silvan Schweber (1994) -- but was interested in the physical structure of this world. The discovery of the 2.73 K background radiation (in 1965) came just in time to trigger my interest in 
Cosmology. The 1971 `Survey of Cosmology' summarizes my understanding at a time when I had not yet learned how many unrealistic assumptions one can make in a field remote from experimental verification.

(6) The global structure of gravitational fields can be non-trivial, as evidenced by the existence of black-hole spacetimes in General Relativity, and of the `big-bang' models. In 1971, an invitation by the Canadian Mathematical Congress gave me an opportunity 
to present a semi-popular survey of this rich field which had been developed successively by Roger Penrose, Steven Hawking, Brandon Carter, Bob Geroch, Hans-J\"urgen Seifert, and by many others.

(7) Starting with a proposal in 1969, and ending with a final report by Eckhard Krotscheck \& myself in 1983, I was principal investigator of experiment 11 on the German-American spaceprobe HELIOS, attempting to test Einstein's theory at the 1\% level. To 
this end, the eccentric orbit of HELIOS (in 3-space) had to be evaluated with a relative in-plane accuracy of $10^{-8.5}$ -- monitored via range ($\ga$ 3m) and range-rate ($\ga 10^{-2}$cm/s) -- whereby all non-gravitational perturbations had to be modelled 
at the same level. (Earlier experiments with a similar nominal precision had been much less reliable). For the evaluation, Otto B\"ohringer and Eckhard Krotscheck prepared the iterated extended batch filter algorithm `COSMOS' which was able to handle a 
heavy n-body problem with several parametrized non-gravitational perturbations. Unfortunately, both missions were sacrificed to the 10 active experiments onboard because their first amplifier tubes burned out when switched to range measurements.

\newpage

\mbox{}

\vspace{2cm}

\noindent {\bf ALTERNATIVES}

\setlength{\leftmargin}{-1cm}

\begin{table}

{\scriptsize
\begin{tabular}{|r|l|l|}
\hline
Nr. &  Alternative & Year \\
\hline
1. & Quakes of neutron-star crusts \{can / cannot\} be treated like those of stressed terrestrial solids. &  75 \\
\hline
2. & The Entropy of a young Black Hole is \{$\approx  / \ll$ \} Hawking's expression;  (entropy = another `hair'). & 76 \\
\hline
3. & The planet Venus \{is / is not\} spin-phase locked to Earth;  (deviation from synchronous $> 10^{-5}$). & 77\\
\hline
4. & The Speed of a Signal \{can / cannot\} exceed the speed of light;  (`front speed' counts). & 78 \\
\hline
5. &  Hydrogen has a (second) critical point at  (p,T) = ($10^{5.38} bar$, $10^{4.28} K$). & 83 \\
\hline
6. & Astrophysical Jets are \{hard / soft\} beams, like a \{lawn sprinkler / hair drier\}. & 80,96 \\
\hline
7.& The Beam (bulk) Velocity is \{$\la$ relativistic / extremely relativistic\}. & 80,96 \\
\hline
 8. & Astrophysical Jets consist of \{hydrogen / (relativistic) pair plasma\}. &  80,96 \\
\hline
 9. & The Beaming pattern in the jets is due to \{bulk velocity / spread in tangents\}. & 80,96 \\
\hline
10. & The bright Knots in the lobes \{are not / are\} pressure-confined. & 80 \\
\hline
11. & In-Situ Acceleration in the knots \{is necessary / is ignorable (violates the Second Law)\}. &  80,84 \\
\hline
12. & The outflow region of a Bipolar Flow is a \{windzone / expanding cocoon\}. &  84,87\\
\hline
13.& The jet in 3C 273 is intrinsically \{1-sided / 2-sided\}. & 86 \\
\hline
14. & The motion of the beam particles is \{gasdynamic / field-guided (cold beam)\}. & 87,89\\
\hline
15. & The Central Engine of an AGN is a \{BH / (SMC or) BD\},  \{black hole / burning disk\}. &79,96 \\
\hline
16. & The BLR of QSOs is filled with \{($10^8 K$) hot hydrogen / pair plasma\}.& 79,87\\
\hline
17. & The Big Blue Bump (UV-Source) in AGN is the \{BH disk / BD\};  (BD = burning disk).& 87,96\\
\hline
18.&The motion in the BLR - and in YY Orionis stars - is \{infall or rotation / outflow\}; (inv. P-Cygni).&88,96\\
\hline
19.&The moving emission lines of SS 433 {are / are not} emitted by the jets; (`bullets' versus e$^{\pm}$ jets). & 79,98\\
\hline
20.&The mapped jets of SS 433 consist of \{local galactic matter / pair plasma\}.& 81,85\\
\hline
21. & The Ly$\alpha$ forest (of QSO absorption lines) is realized by \{static clouds / ejected filaments\}. & 85\\
\hline
22. & Galaxies in clusters evolve by \{merging (or stripping) / harassing\}; [A. Toomre].& 85\\
\hline
23. & The Wisps in the Crab are emitted \{incoherently / coherently (i.e. are a LASER)\}. & 77\\
\hline
24. & The Crab PSR wind \{is / is not\} strong enough to post-accelerate the filaments (by some 8\%). & 80,90\\
\hline
25.& Cosmic Rays are accelerated by \{(interstellar) shocks / neutron stars\}. & 78,98\\
\hline
26. & The highest-energy Cosmic Rays ($> 10^{19} eV$) are of \{extragalactic / Galactic\} origin.& 89\\
\hline
27. & The Black-Hole candidates involve \{black holes / neutron stars (with massive disks)\}. &  79,89\\
\hline
28.& Wolf-Rayet stellar winds \{are / are not\} radiatively driven. & 90\\
\hline
29. & (Split, broad) Emission Lines of compact sources come from their \{accretion disk / windzone\}. &  89,96\\
\hline
30. & Neutron-star dipole moments \{do / do not\} decay (within $10^{10} yr$). &  81,94\\
\hline
31.&Pulsar winds consist of pair plasma, post-accelerated by \{certain fields / strong outgoing wave\}.& 86,98\\
\hline
32.& Pair-plasma winds, or jets, are generated by \{exceptional / all\} neutron stars. & 86,98\\
\hline
33.& Neutron stars form from (evolved) stars with  M $\ga \{M_{\odot}$ (e.g. white dwarf) / 3 $M_{\odot}$\}.&  85,98\\
\hline
34.& Neutron stars derive from \{often single / mostly binary\} stars.& 85,98\\
\hline
35.& Pulsar beams are \{pencil / fan\} beams; i.e. PSRs have a \{small / large\} beaming fraction. &  85,98\\
\hline
36.& Pulsar radio pulses come from \{near polar caps / inside the speed-of-light cylinder\}. & 85\\
\hline
37.& Accreting X-ray sources are \{sometimes wind-fed / always disk-fed\}.& 85\\
\hline
38.& A common envelope \{can / cannot\} form around a neutron star. &  85\\
\hline
39.&The msec pulsars are \{spun up by accretion (`recycled') / born fast\}.&85,98\\
\hline
\end{tabular}
}
\end{table}

\newpage


\mbox{}

\vspace{2cm}

\begin{table}
{\scriptsize
\begin{tabular}{|r|l|l|}
\hline
40.& Neutron-star accretion takes place onto \{polar caps, along $\vec B$ / equatorial belt (also), as `blades'\}.& 87,98\\
\hline
41.& The non-pulsing n-stars and msec PSRs have \{weak ($< 10^{10} G$) / strong ($> 10^{11} G$)\} surf. magn. fields.&  87,98\\
\hline
42.& X-ray QPOs stem from \{accretion flow / corotating magnetosphere (near speed-of-light cylinder)\}. & 89,98\\
\hline
43.& The (reported) pulsar in SN 1987A has a (spin) period \{ $<$  / $>$ \} 1 msec;  [printed at critical epoch].& 90\\
\hline
44.& Pulsar torque noise and glitches are due to changes in \{moment of inertia / (also) dipole moment\}.&94,98\\
\hline
45.& Pulsar radio emission is \{ill-understood / a MAFER (= Microw. Amplifier by Forced Em. of  Radiation)\}.  & 95,98\\
\hline
46.& Pulsar proper motions \{can exceed $10^3 Kms^{-1}$/ are  $\la 10^{2.7} Kms^{-1}$\}.& 95,98\\
\hline
47.& Atmospheric Superrotation on the Sun and planets is driven \{internally / magnetically or externally\}.& 83,98\\
\hline
48.& The solar magnetic flux is \{generated / modulated\} by its convection zone; (\{is not / is\} anchored in core).& 92,93\\
\hline
49. & The High-Velocity Clouds in the upper Galactic hemisphere \{do not map / map\} the Galactic Jet. & 87,92\\
\hline
50.& Sgr A West lies \{in front / inside\} of Sgr A East;  the latter \{is / is not\} a SN remnant. & 90,96\\
\hline
51.& The mass of Sgr A* is \{$\ga 10^6 M_{\odot}$ / $\la 10^3 M_{\odot}$\}; (Sgr A* = radio-point source at center of Galaxy).& 90,96\\
\hline
52.& The Supernova piston consists of \{neutrinos / magnetic torsional spring plus pair plasma\}.& 76,88\\
\hline
53.& Supernova explosions \{can / cannot (almost always)\} give birth to black holes.& 85,98\\
\hline
54. &Supernovae behave like \{pressure / splinter (shrapnel)\} bombs;  (i.e. are \{thin-walled / thick-walled \}).&  88,98\\
\hline
55.& Supernova Shells receive their relativistic electrons (and positrons) \{in situ / at birth\}.&  88,90\\
\hline
56.& Supernova Shells are \{multiple shock waves / flaring former windzones (traversed by filaments)\}.&85,95\\
\hline
57.& Supernova shells lose their kinetic energy to \{radiation / galactic-disk expansion\}.&  88\\
\hline
58.& Supernova light curves are powered by \{radioactive decay / explosion energy plus e$^{\pm}$ plus n-star cooling\}. & 88,98\\
\hline
59.& The `exotic' Supernova Remnants are \{multiple events / Pulsar Nebulae\}; (e.g. CTB 80). &  92,98\\
\hline
60.& The Fireworks in Orion are a \{bipolar flow / (young) supernova remnant\}.&   95\\
\hline
61.& The `fossil' fuels (natural gas, oil, and coal) are of \{biogenic / abiogenic\} origin; [W.Plotts, T.Gold].&  86\\
\hline
62.& Plate tectonics on Earth are driven by \{thermal mantle currents / volcanic `fences', rooted in core\}.& 86,91\\
\hline
63.& The mantle of Earth is a \{good / poor\} conductor;  (i.e. magnetically permeable). &  89\\
\hline
64.& The LOD decadic fluctuations are due to \{core-mantle coupling / atmospheric spin and/or changes of I\}.&  89\\
\hline
65.& The Galactic Disk is filled with \{(mainly) hydrogen / pair plasma\}, escaping through `chimneys'. & 87,92\\
\hline
66. &Accretion Disk dynamics is controlled by \{turbulence / toroidal magnetic fields\}. & 90\\
\hline
67. & The (daily) $\gamma$-ray Bursts come from \{$\ge$ halo distances / nearby n-stars\} (sparks above n-star surface).&93\\
\hline
68. &The soft $\gamma$-ray Repeaters are at \{halo / nearby ($\la 50 pc$, inside pulsar nebula)\} distances.&  94\\
\hline
69.& The 2.73 K background radiation owes its blackness to \{fast decoupling / hydrogen snow\}; (?). & 95\\
\hline
70.& The oldest (Galactic) stars are  \{$\ge 14 Gyr$/$\ga 10 Gyr$\} old.&  96\\
\hline
71.& Superhump periods in X-ray binaries are produced by \{disk disturbances / extra-Roche clumps\}. & 96,98\\
\hline
72.& The (bright, unresolved) supersoft X-ray sources (SSS) are powered by \{white dwarfs / neutron  stars\}.&  96,98\\
\hline
73.& Spermwhales dive to the bottom of the Sea \{with / without\} having to perform work.&  92\\
\hline
74.& The Na-K Pumps in cell membranes act on the \{K / Na\}-ions, as heat pumps (generating $\approx 0.07$ V).&  94\\
\hline
75.& (Certain) Birds can sense magnetic fields \{directly / electrically (via the Lorentz force)\}.&  95\\
\hline
76.& The rise of water in Plants is \{not / often\} achieved by single-cell mechanical pumps (in root tips).&96,98\\
\hline
77.& Photosynthesis in plants involves \{electron / proton\} currents;  (the e's cascade through bound states).&96\\
\hline
78.& The heavenly hail `detected' by L. Frank via UV holes \{is / is not\} realized by house-sized snowballs.&  87,97\\
\hline
79.& The Earth's atmosphere \{is / is not\} charged by its thunderstorm clouds.&  98\\
\hline

\end{tabular}
}
\end{table}


\end{document}